\documentclass[final,3p]{elsarticle}

 \usepackage{graphicx}

\usepackage{amssymb}

\usepackage{amsmath}
\usepackage{dsfont}
\usepackage{bm}
\usepackage{mathrsfs}

\begin{document}

\begin{frontmatter}



\title{Casimir stress in an inhomogeneous medium}


\author[phys]{T.G.~Philbin}
\ead{tgp3@st-andrews.ac.uk}
\author[com]{C.~Xiong}
\author[phys]{U.~Leonhardt}

\address[phys]{School of Physics and Astronomy,  University of St Andrews, North Haugh, St Andrews. Fife KY16~9SS, Scotland}
\address[com]{School of Computer Science,  University of St Andrews, North Haugh, St Andrews. Fife KY16~9SX, Scotland}

\begin{abstract}
The Casimir effect in an inhomogeneous dielectric is investigated using Lifshitz's theory of electromagnetic vacuum energy. A permittivity function that depends continuously on one Cartesian coordinate is chosen, bounded on each side by homogeneous dielectrics. The result for the Casimir stress is infinite everywhere inside the inhomogeneous region, a divergence that does not occur for piece-wise homogeneous dielectrics with planar boundaries. A Casimir force per unit volume can be extracted from the infinite stress but it diverges on the boundaries between the inhomogeneous medium and the homogeneous dielectrics. An alternative regularization of the vacuum stress is considered that removes the contribution of the inhomogeneity over small distances, where macroscopic electromagnetism is invalid. The alternative regularization yields a finite Casimir stress inside the inhomogeneous region, but the stress and force per unit volume diverge on the boundaries with the homogeneous dielectrics. The case of inhomogeneous dielectrics with planar boundaries thus falls outside the current understanding of the Casimir effect. 
\end{abstract}


\end{frontmatter}



\section{Introduction}
Quantization of the electromagnetic field leads to a vacuum, zero-point energy-momentum that diverges~\cite{milonni}. As this infinite vacuum energy clearly does not exist it was initially assumed that the infinity should be zero, until Casimir showed~\cite{cas48} that a finite vacuum energy can sometimes be extracted, leading to the prediction of measurable forces.

In recent years there has been considerable progress in the experimental demonstration of Casimir forces between objects~\cite{bor09}, and a growing appreciation of the problems and opportunities these forces entail for micro- and nano-engineering~\cite{nature,cap07}. On the theory side, it is understood how to calculate the Casimir forces between an arbitrary number of separated bodies, the most versatile approach being that of Lifshitz~\cite{lif55,dzy61,LL} wherein the electromagnetic stress tensor and energy density are obtained from the Green tensor for the electric field (or vector potential). Although the calculations for any arrangement of objects beyond the classic case of two parallel half-spaces~\cite{cas48,lif55} are extraordinarily cumbersome, the problem can be solved purely numerically using standard codes for the Green tensor~\cite{rod09,rei09,mcc09}. 

Yet the theoretical understanding of the Casimir effect remains incomplete in some fundamental respects.  Most importantly, the Casimir self-energy of an object is a quantity whose exact meaning is still debated and for which the standard approaches like Lifshitz theory give a diverging answer in general~\cite{bor09}. The subject of Casimir self-energy began in 1968 with Boyer's spectacular conclusion that the self-energy of an infinitesimally thin, perfectly conducting spherical shell is positive, giving a repulsive (i.e.\ outwardly directed) Casimir force on the shell~\cite{boy68}. An unsettling feature of the spherical shell is the occurrence of an additional divergence which is not present in the case of perfectly conducting parallel plates~\cite{cas48} and which must be regularized to obtain a finite answer. It might be hoped that the extra divergence is due to the idealized properties of vanishing thickness and perfect conduction of the shell, but the apparently reasonable case of spherically symmetric materials with general electric permittivity and magnetic permeability is even more ill-behaved. Even with allowance for (temporal) dispersion, the additional divergences remain and there is no agreement on their precise significance, or how to remove them~\cite{mil78,deu79,mil01,gra04,bar04a,bar04b,mil06,bor08,bor09}. Hence the striking fact that the seemingly innocuous issue of the Casimir self-energy of a dielectric ball is still not understood~\cite{mil01,bor09}. Similar problems occur for cylindrically symmetric media, the other case where Casimir self-forces have been investigated in some detail. (See~\cite{bor09}  for a guide to the Casimir literature on spherical and cylindrical geometries.) This lack of understanding of the self-force for these geometries makes any attempt to compute the Casimir forces on concentric spherical and cylindrical bodies problematic.

The work described above has all been for piece-wise homogeneous media, with the non-zero Casimir forces arising from the boundaries between the homogeneous components. Although the general theory of Lifshitz and his co-workers~\cite{lif55,dzy61,LL} is formulated for inhomogeneous, dispersive dielectrics,  to our knowledge there is no example in the literature where the Casimir effect for an inhomogeneous dielectric is actually calculated.\footnote{A previous study~\cite{inu08} considered layers of homogeneous slabs as an approximation to an inhomogeneous dielectric and computed the Casimir energy arising from the TM evanescent modes of the system.}  In this paper we provide such an example by using Lifshitz theory to calculate the vacuum electromagnetic stress tensor for a simple model of an inhomogeneous dielectric. The stress tensor specifies the local Casimir self-force in the medium arising from its inhomogeneity. We choose the simplest example where the inhomogeneity is in one spatial dimension only, the medium being homogeneous in planes orthogonal to this direction. Boundaries are imposed on both sides of the inhomogeneous medium, beyond which the material is homogeneous. As the boundaries are planar, we expected to avoid the problems with divergences described above, which appear to be associated with curved boundaries and a high degree of symmetry. Our results did not meet these expectations. Specifically, we find that the regularization prescription for inhomogeneous dielectrics advocated in standard Lifshitz theory~\cite{dzy61,LL} yields an infinite Casimir stress everywhere inside the inhomogeneous region. A Casimir force per unit volume, given by the spatial derivative of the stress, can be extracted; this appears to be finite inside the inhomogeneous region but it diverges on the boundaries with the homogeneous dielectrics. We investigate these divergences by considering a new regularization procedure, based on a consideration of the limited validity of macroscopic electromagnetism. The new regularization gives a finite Casimir stress inside the inhomogeneous medium, however the stress and the force per unit volume diverge on the boundaries with the homogeneous regions. We thus find that, in addition to spherically and cylindrically symmetric dielectrics with boundaries, inhomogeneous dielectrics with planar boundaries contain divergences that fall outside current understanding of the Casimir effect.

In Section~\ref{theory} we review Lifshitz theory for a medium inhomogeneous in one spatial dimension. The problem of calculating the Casimir stress reduces to solving the equations for two scalar Green functions, associated with the TE (transverse electric) and TM (transverse magnetic) modes. We discuss the problem of regularizing the quantum-vacuum energy-momentum to obtain a finite Casimir stress. In addition to the standard regularization we consider an alternative that removes the contribution to the Casimir stress of the inhomogeneity over small distances where macroscopic electromagnetism is invalid. In Section~\ref{example} we consider a particular functional form for the permittivity of an inhomogeneous dielectric bounded by homogeneous regions and obtain analytic solutions for the Green functions. The resulting Casimir stress is considered using both the standard and the alternative regularizations; we find that neither gives a Casimir stress and force per unit volume that are everywhere finite.

\section{Lifshitz theory for media inhomogeneous in one Cartesian coordinate}  \label{theory}
In the formalism of Lifshitz~\cite{lif55,dzy61,LL} the forces on macroscopic bodies due to the electromagnetic quantum vacuum are obtained by calculating the vacuum expectation value of the Maxwell stress tensor:
\begin{equation} \label{stress}
\bm{\sigma}=\langle \mathbf{\hat{D}}\otimes\mathbf{\hat{E}}\rangle+\langle \mathbf{\hat{H}}\otimes\mathbf{\hat{B}}\rangle
-\frac{1}{2}\mathds{1}\left(\langle \mathbf{\hat{D}}\cdot\mathbf{\hat{E}}\rangle+\langle \mathbf{\hat{H}}\cdot\mathbf{\hat{B}}\rangle\right).
\end{equation}
The materials are in general inhomogeneous and dispersive with dielectric permittivity $\varepsilon(\mathbf{r},\omega)$ and magnetic permeabilty $\mu(\mathbf{r},\omega)$; anisotropy may also be included but here we consider only scalar $\varepsilon$ and $\mu$
The vacuum correlation functions of electric and magnetic fields in (\ref{stress}) are given by
\begin{gather} 
\langle\mathbf{\hat{D}}(\mathbf{r},t)\!\otimes\!\mathbf{\hat{E}}(\mathbf{r},t)\rangle=-\frac{\hbar}{\pi c^2}\int_0^\infty d\xi\,\varepsilon(\mathbf{r},i \xi)\,\xi^2\mathbf{G}(\mathbf{r},\mathbf{r},i \xi),   \label{EE} \\[5pt]
\langle\mathbf{\hat{H}}(\mathbf{r},t)\!\otimes\!\mathbf{\hat{B}}(\mathbf{r},t)\rangle
=\frac{\hbar}{\pi}\lim_{\mathbf{r'}\rightarrow\mathbf{r}}\int_0^\infty d\xi\,\frac{1}{\mu(\mathbf{r},i \xi)}\nabla\times\mathbf{G}(\mathbf{r},\mathbf{r'},i \xi)\times\stackrel{\leftarrow}{\nabla'}.  \label{BB}
\end{gather}
In (\ref{EE})--(\ref{BB}), $\mathbf{G}(\mathbf{r},\mathbf{r'},i \xi)$ is the imaginary-frequency retarded Green tensor for the vector potential in a gauge in which the electric potential is set to zero; its equation is
\begin{equation} \label{green}
\left(\nabla\times\frac{1}{\mu(\mathbf{r},i \xi)}\nabla\times+\frac{\xi^2}{c^2}\varepsilon(\mathbf{r},i \xi)\right)\mathbf{G}(\mathbf{r},\mathbf{r'},i \xi)=\mathds{1}\delta(\mathbf{r}-\mathbf{r'}).
\end{equation}
The notation $\times\stackrel{\leftarrow}{\nabla'}$ in (\ref{BB}) denotes a curl on the second index of $\mathbf{G}(\mathbf{r},\mathbf{r'},i \xi)$,  so that for a vector $\mathbf{V}(\mathbf{r'})$ we have $\mathbf{V}\times\stackrel{\leftarrow}{\nabla'}=\nabla'\times\mathbf{V}$. Physically, (\ref{green}) represents a dipole oscillating with frequency $i \xi$ at the point $\mathbf{r'}$ and $\mathbf{G}(\mathbf{r},\mathbf{r'},i\xi)$ is the resulting vector potential at the point $\mathbf{r}$. The second index in $G_{ij}$ represents the orientation of the dipole at $\mathbf{r'}$, while the first index represents the components of the vector potential at $\mathbf{r}$. An important property of the Green tensor, which holds for systems invariant under time-reversal~\cite{LL}, is
\begin{equation} \label{recip}
G_{ij}(\mathbf{r},\mathbf{r'},i \xi)=G_{ji}(\mathbf{r'},\mathbf{r},i \xi).
\end{equation}
The Casimir force $\mathbf{F}$ on a body occupying a volume $\Omega$ is obtained by integrating the stress tensor (\ref{stress}) over the boundary surface $\partial\Omega$ of the body:
\begin{equation}  \label{F}
\mathbf{F}=\int_{\partial \Omega}\bm{\sigma}\cdot d\mathbf{S}=\int_ \Omega\nabla\cdot\bm{\sigma}\,dV.
\end{equation}
An infinitesimal element $dV$ of the body at position $\mathbf{r}$ experiences a local force $\mathbf{f}\,dV$, where $\mathbf{f}$ is the force per unit volume at $\mathbf{r}$:
\begin{equation}   \label{f}
\mathbf{f}=\nabla\cdot\bm{\sigma}.
\end{equation}
 
The stress tensor (\ref{stress}) is thus sufficient to determine the Casimir forces on the materials, but one can also calculate the quantum-vacuum energy density. Since the materials are dispersive, this energy density is given by Brillouin's formula~\cite{LLcm}. In terms of the monochromatic (in imaginary frequency) expectation values
\begin{gather} 
\langle\mathbf{\hat{E}}(\mathbf{r})\!\otimes\!\mathbf{\hat{E}}(\mathbf{r})\rangle_\xi=-\frac{\hbar\mu_0}{\pi}\xi^2\mathbf{G}(\mathbf{r},\mathbf{r},i \xi),   \label{EE2} \\[5pt]
\langle\mathbf{\hat{H}}(\mathbf{r})\!\otimes\!\mathbf{\hat{H}}(\mathbf{r})\rangle_\xi
=\frac{\hbar}{\pi\mu_0}\lim_{\mathbf{r'}\rightarrow\mathbf{r}}\frac{1}{\mu^2(\mathbf{r},i \xi)}\nabla\times\mathbf{G}(\mathbf{r},\mathbf{r'},i \xi)\times\stackrel{\leftarrow}{\nabla'}, \label{BB2}
\end{gather}
the Casimir energy density is
\begin{equation}  \label{energy}
\rho=\frac{1}{2}\int_0^\infty d\xi\left(\varepsilon_0\frac{d(\xi\varepsilon(\mathbf{r},i \xi))}{d\xi}\langle \mathbf{\hat{E}}\cdot\mathbf{\hat{E}}\rangle_\xi+\mu_0\frac{d(\xi\mu(\mathbf{r},i \xi))}{d\xi}\langle \mathbf{\hat{H}}\cdot\mathbf{\hat{H}}\rangle_\xi\right).
\end{equation}

The Casimir stress (\ref{stress}) and energy density (\ref{energy}) are thus obtained by solving (\ref{green}) for the Green tensor, subject to the boundary conditions imposed by the specific materials. A drawback to this approach is that the number of cases where (\ref{green}) can be solved analytically with non-trivial boundary conditions is severely limited. On the other hand, the task of solving (\ref{green}) for a given arrangement of objects is a well-posed classical
electromagnetism problem. It is the issue of regularizing the Green tensor so that it yields a finite Casmir force that sets the current limit to standard Lifshitz theory, since it is not understood how to do this in general. As described in the Introduction, the Casimir effect for general dielectrics with spherical and cylindrical symmetry is still unsolved, even though the Green tensor in each case is known. Before discussing the regularization problem for inhomogeneous media, we show how to solve (\ref{green}) in an inhomogeneous medium where  $\varepsilon(\mathbf{r},\omega)$ and $\mu(\mathbf{r},\omega)$ depend only on one Cartesian coordinate~\cite{sch77,leo}.

\subsection{The Green tensor}
Let $\varepsilon(\mathbf{r},\omega)=\varepsilon(x,\omega)$ and $\mu(\mathbf{r},\omega)=\mu(x,\omega)$ depend only on the $x$-coordinate. The resulting homogeneity in the $yz$-plane allows us to Fourier transform the Green tensor:
\begin{equation} \label{four}
\mathbf{\widetilde{G}}(x,x',u,v,i \xi) 
=\int_{-\infty}^\infty d y\int_{-\infty}^\infty d z\,\mathbf{G}(\mathbf{r},\mathbf{r'},i \xi)\,e^{-i u(y-y')-i v(z-z')},
\end{equation}
so that the radiation emitted by the dipole is decomposed into waves with wave-vector components $k_y=u$, $k_z=v$ at the imaginary frequency $i\xi$.
With these symmetry assumptions, we are restricted to considering planar material
 boundaries, given by values of $x$. This is one of the cases where the vector-potential waves (proportional to the electric field) can be decomposed into independent TE (electric field in $yz$-plane) and TM (magnetic field in $yz$-plane) waves. The general solution for the Fourier-transformed Green tensor (\ref{four}) can then be written in terms of \textit{scalar} Green functions $\widetilde{g}_E(x,x',u,v,i \xi)$ and $\widetilde{g}_M(x,x',u,v,i \xi)$ for the TE and TM waves~\cite{sch77,leo}. 

To write the equation for the Fourier-transformed Green tensor (\ref{four}) we introduce the vector operators $\boldsymbol{\mathcal{D}}$ and  $\boldsymbol{\mathcal{D}'}$, which function as Fourier transforms of the operators $\nabla$ and $\nabla'$:
\begin{equation}  \label{Ddef}
\boldsymbol{\mathcal{D}}:=(\partial_x,i u,i v), \qquad \boldsymbol{\mathcal{D}'}:=(\partial'_x,-i u,-i v).
\end{equation}
We then obtain from (\ref{green}) and (\ref{four})
\begin{equation} \label{foureqn}
\left(\boldsymbol{\mathcal{D}}\times\frac{1}{\mu(x,i \xi)}\boldsymbol{\mathcal{D}}\times+\frac{\xi^2}{c^2}\varepsilon(x,i \xi)\right)\mathbf{\widetilde{G}}(x,x',u,v,i \xi) =\mathds{1}\delta(x-x'),
\end{equation}
and the solution of this equation has the reciprocity property that follows from (\ref{recip}) and (\ref{four}): 
\begin{equation} \label{recipfour}
\widetilde{G}_{ij}(x,x',u,v,i \xi)=\widetilde{G}_{ji}(x',x,-u,-v,i \xi).
\end{equation}
The scalar Green functions $\widetilde{g}_E$ and $\widetilde{g}_M$ are defined by
\begin{eqnarray}
\left(\boldsymbol{\mathcal{D}}\cdot\frac{1}{\mu(x,i \xi)}\boldsymbol{\mathcal{D}}-\frac{\xi^2}{c^2}\varepsilon(x,i \xi)\right)\widetilde{g}_E(x,x',u,v,i \xi) =\delta(x-x'),  \label{gE} \\[5pt]
\left(\boldsymbol{\mathcal{D}}\cdot\frac{1}{\varepsilon(x,i \xi)}\boldsymbol{\mathcal{D}}-\frac{\xi^2}{c^2} \mu(x,i \xi)\right)\widetilde{g}_M(x,x',u,v,i \xi) =\delta(x-x'),  \label{gM}
\end{eqnarray}
and exhibit the same reciprocity symmetry (\ref{recipfour}):
\begin{eqnarray}
\widetilde{g}_E(x,x',u,v,i \xi) =\widetilde{g}_E(x',x,-u,-v,i \xi),  \label{recipE} \\
\widetilde{g}_M(x,x',u,v,i \xi) =\widetilde{g}_M(x',x,-u,-v,i \xi)  \label{recipM} 
\end{eqnarray}
We also define a unit vector in the direction of the TE electric field:
\begin{equation}  \label{nE}
\mathbf{n}_{E}:=\frac{1}{\sqrt{u^2+v^2}}\left(\begin{array}{c} 0 \\ -v \\ u \end{array}\right).
\end{equation}
The general solution of (\ref{foureqn}) for the Fourier-transformed Green tensor $\mathbf{\widetilde{G}}$ can now be written as
\begin{equation} 
\mathbf{\widetilde{G}}=-\mathbf{n}_{E}\,\widetilde{g}_E\otimes\mathbf{n}_{E}+c^2\frac{\boldsymbol{\mathcal{D}}\times\mathbf{n}_{E}\,\widetilde{g}_M\otimes\mathbf{n}_{E}\times\stackrel{\leftarrow}{\boldsymbol{\mathcal{D}}'}}{\varepsilon(x,i \xi)\,\varepsilon(x',i \xi)\,\xi^2} 
+\frac{c^2}{\varepsilon(x',i \xi)\,\xi^2}\mathds{1}\delta(x-x')-\frac{c^2}{\varepsilon(x',i \xi)\,\xi^2}\mathbf{n}_{E}\otimes\mathbf{n}_{E}\,\delta(x-x'). \label{Gsol}
\end{equation}

To confirm (\ref{Gsol}) as a solution of (\ref{foureqn}) requires the following identities
for an arbitrary function $f(x)$ and vector $\mathbf{V}(x)$:
\begin{gather}
\left[\boldsymbol{\mathcal{D}}\times f(x)\boldsymbol{\mathcal{D}}\times\mathbf{V}(x)\right]_i=\mathcal{D}_jf(x)\mathcal{D}_iV_j(x)-\mathcal{D}_jf(x)\mathcal{D}_jV_i(x).  \label{id1}  \\
\boldsymbol{\mathcal{D}}\cdot\mathbf{n}_{E}f(x)=0.  \label{id2}
\end{gather}
The effect of  the differential operator in (\ref{foureqn}) on the terms in  (\ref{Gsol}) containing the scalar Green functions is found from the following, which are obtained by successively applying (\ref{id1}), (\ref{id2}) and the relevant Green-function equation (\ref{gE}) or (\ref{gM}):
\begin{eqnarray}
\boldsymbol{\mathcal{D}}\times\frac{1}{\mu(x,i \xi)}\boldsymbol{\mathcal{D}}\times\mathbf{n}_{E}\,\widetilde{g}_E=-\frac{\xi^2}{c^2}\varepsilon(x,i \xi)\,\mathbf{n}_{E}\,\widetilde{g}_E-\mathbf{n}_{E}\,\delta(x-x'),  \label{DDgE} \\[5pt]
\frac{1}{\mu(x,i \xi)}\boldsymbol{\mathcal{D}}\times\frac{\boldsymbol{\mathcal{D}}\times\mathbf{n}_{E}\,\widetilde{g}_M}{\varepsilon(x,i \xi)}=-\frac{\xi^2}{c^2}\,\mathbf{n}_{E}\,\widetilde{g}_M-\frac{1}{\mu(x,i \xi)}\mathbf{n}_{E}\,\delta(x-x').   \label{DDgM}
\end{eqnarray}
When (\ref{Gsol}) is inserted into the left-hand side of (\ref{foureqn}) and (\ref{DDgE})--(\ref{DDgM}) are employed, one obtains only terms containing $\delta(x-x')$; the latter simplify to the right-hand side of (\ref{foureqn}) because of the identity
\begin{equation}
\boldsymbol{\mathcal{D}}\times\frac{1}{\mu(x,i \xi)}\boldsymbol{\mathcal{D}}\times\left[\mathds{1}\delta(x-x')-\mathbf{n}_{E}\otimes\mathbf{n}_{E}\,\delta(x-x')\right]   
+\boldsymbol{\mathcal{D}}\times\frac{1}{\mu(x,i \xi)}\mathbf{n}_{E}\otimes\mathbf{n}_{E}\,\delta(x-x')\times\stackrel{\leftarrow}{\boldsymbol{\mathcal{D}}'}=0,
\end{equation}
which must be verified using (\ref{id1}) and the definitions (\ref{Ddef}) and (\ref{nE}).

The problem of finding the Green tensor is thus reduced to that of solving the two ordinary differential equations (\ref{gE}) and (\ref{gM}) for the scalar Green functions.  The conditions that $\widetilde{g}_E$ and $\widetilde{g}_M$ must obey at material boundaries, values of $x$ where $\varepsilon$, $\mu$ or their derivatives are discontinuous, are obtained from  (\ref{gE}) and (\ref{gM}) in the same manner that the boundary conditions for electromagnetic fields are obtained from Maxwell's equations~\cite{jac}. The boundary conditions are continuity of
\begin{equation}  \label{boundary}
\widetilde{g}_E, \qquad \widetilde{g}_M, \qquad \frac{1}{\mu(x,i \xi)}\partial_x\widetilde{g}_E,  \qquad\mbox{and}\qquad \frac{1}{\varepsilon(x,i \xi)}\partial_x\widetilde{g}_M.
\end{equation}
The continuity of (\ref{boundary}) ensures the correct boundary conditions on the Green tensor (\ref{Gsol}), which are familiar from basic electromagnetism~\cite{jac} since $\mathbf{G}(\mathbf{r},\mathbf{r'},i\xi)$, $\varepsilon\mathbf{G}(\mathbf{r},\mathbf{r'},i\xi)$, $\nabla\times\mathbf{G}(\mathbf{r},\mathbf{r'},i\xi)$, and $\mu^{-1}\nabla\times\mathbf{G}(\mathbf{r},\mathbf{r'},i\xi)$ are proportional to $\mathbf{E}(\mathbf{r},i\xi)$, $\mathbf{D}(\mathbf{r},i\xi)$, $\mathbf{B}(\mathbf{r},i\xi)$ and $\mathbf{H}(\mathbf{r},i\xi)$, respectively.

Written out explicitly using (\ref{Ddef}), the equations (\ref{gE})--(\ref{gM}) for the scalar Green functions are
\begin{gather}
\left(\partial_x\frac{1}{\mu(x,i \xi)}\partial_x-\frac{u^2+v^2}{\mu(x,i \xi)}-\frac{\xi^2}{c^2} \varepsilon(x,i \xi)\right)\widetilde{g}_E(x,x',u,v,i \xi) =\delta(x-x'), \label{gEsim} \\[5pt]
\left(\partial_x\frac{1}{\varepsilon(x,i \xi)}\partial_x-\frac{u^2+v^2}{\varepsilon(x,i \xi)}-\frac{\xi^2}{c^2} \mu(x,i \xi)\right)\widetilde{g}_M(x,x',u,v,i \xi) =\delta(x-x').  \label{gMsim}
\end{gather}
In a homogeneous region, where $\varepsilon$ and $\mu$ in (\ref{gEsim})--(\ref{gMsim}) are independent of $x$, the general solutions are
\begin{gather}
\widetilde{g}_{\text{h}E} =\widetilde{g}_{0\text{h}E} +c_{\text{h}E1}e^{wx}+c_{\text{h}E2}e^{-wx},    \qquad
\widetilde{g}_{\text{h}M} =\widetilde{g}_{0\text{h}M} +c_{\text{h}M1}e^{wx}+c_{\text{h}M2}e^{-wx},  \label{ghomtot}   \\[5pt]
w:=\sqrt{u^2+v^2+\varepsilon(i \xi)\mu(i \xi)\frac{\xi^2}{c^2}},  \label{wdef}
\end{gather}
where the $c$s are arbitrary constants and $\widetilde{g}_{0\text{h}E} $,  $\widetilde{g}_{0\text{h}M} $ are the (retarded) Green functions for a space-filling homogeneous medium:
\begin{equation}
\widetilde{g}_{0\text{h}E} =-\frac{\mu(i \xi)}{2w}\,e^{-w|x-x'|},   \qquad
\widetilde{g}_{0\text{h}M} =-\frac{\varepsilon(i \xi)}{2w}\,e^{-w|x-x'|}.  \label{ghom} 
\end{equation}

In the next section we will solve (\ref{gEsim})--(\ref{gMsim}) analytically for an inhomogeneous medium. First we must consider the problem of regularizing the Green functions, since use of the full Green tensor in (\ref{EE}) and (\ref{BB}) always yields an infinite Casimir stress (\ref{stress}).

\subsection{Regularization and Casimir stress} 
In the formalism developed by Lifshitz and coworkers~\cite{lif55,dzy61,LL} for realistic dielectrics, the only meaningful quantum-vacuum energy is viewed as that which arises from material inhomogeneity, through smooth or sharp spatial variations of $\varepsilon$ and $\mu$. Thus, according to this viewpoint, the (infinite) quantum-vacuum energy-momentum of a space-filling homogeneous medium that emerges from the theory must be regularized to zero. This infinite energy-momentum is obtained by using the homogeneous Green functions ({\ref{ghom}) to calculate the Casimir stress (\ref{stress}) and energy density (\ref{energy}); hence, at the level of the Green functions, the regularization prescription is to discard the homogeneous Green functions (\ref{ghom}). 

The simplest case where a meaningful non-zero Casimir energy-momentum arises is that of piece-wise homogenous media with sharp boundaries. Here the regularization prescription is clear from that for a space-filling homogeneous medium. This is because, for piece-wise homogeneous media, the solution for the Green tensor (\ref{green}) in each homogeneous sub-medium is a sum of the \textit{bare} or \textit{bulk} Green tensor, which is the solution in the absence of a boundary to the sub-medium, and the \textit{scattered} Green tensor, which ensures that the electromagnetic boundary conditions are satisfied by the total Green tensor on the boundary of the sub-medium. (For example, this decomposition into bulk and scattered parts is clear in the scalar Green functions (\ref{ghomtot})). From what has been said regarding a space-filling homogeneous medium, it is clear that the bulk Green tensor must always be dropped, since it has been ruled out as a source of Casimir energy. It follows that the Casimir energy-momentum is given by the scattered Green tensor. Unfortunately, the scattered Green tensor does not always yield finite Casimir forces on the materials. The most studied examples where the scattered Green tensor gives an infinite Casimir force are those of spherically and cylindrically symmetric dielectrics. This divergence problem emerged in Boyer's original treatment of an infinitely thin, perfectly conducting shell~\cite{boy68}, where he used a mode summation technique. Within standard Lifshitz theory, any divergence is supposed to arise from the bulk Green tensor only, and there is no prescription for dealing with additional divergences that stem from the scattered Green tensor. Although in a few cases such as Boyer's shell, additional regularizations have been performed to obtain a finite Casimir force, there has been no proposal for a regularization procedure that will work for general dielectrics with spherical and cylindrical symmetry. Moreover, there is no consensus on the exact meaning of the additional divergences or on the validity of regularizations like Boyer's~\cite{mil78,deu79,mil01,gra04,bar04a,bar04b,mil06,bor08,bor09}. Methods other than Lifshitz theory have also been unsuccessful in obtaining a (finite) prediction for the Casimir force on general spherically and cylindrically symmetric dielectrics~\cite{bor09}.

The paragon of a successful calculation of the Casimir force for realistic materials is the Lifshitz-theory formula for the force between parallel half-spaces made of homogeneous dielectrics, separated by a third homogeneous dielectric~\cite{dzy61}. This formula has been tested experimentally, both directly and through its prediction of the approximate force when one of the half-spaces is replaced by a sphere~\cite{bor09}. Here the regularization prescription of dropping the bulk Green tensor yields a finite Casimir force on the half-spaces, but even in this example the issue of divergences is not entirely straightforward.~\footnote{The following results are probably well known to the experts, but they are not usually recorded in presentations of the Lifshitz theory for parallel half-spaces.} If the $x$ axis is perpendicular to the planar boundaries of the half-spaces, then (\ref{F}) shows that the perpendicular force on the half-spaces is determined by $\sigma_{xx}$, which is zero inside the half-spaces but jumps to a constant finite value between them.  Although  $\sigma_{xx}$ is sufficient to find the Casimir force, additional information on the vacuum energy-momentum is provided by the other components of the (diagonal) $\bm{\sigma}$ and by the energy density (\ref{energy}). It turns out that  $\sigma_{yy}$, $\sigma_{zz}$ and $\rho$ all diverge on the surfaces of the half-spaces. The fact that $\rho$ is not finite seems to raise the possibility of a contradiction in the theory, since the force can also be computed by integrating $\rho$ to obtain the Casimir energy and differentiating this with respect to the distance between the half-spaces. If this path to the Casimir force is followed it turns out that the terms causing the divergence of $\rho$ on the boundaries disappear upon differentiation with respect to the distance between the half-spaces, and the resulting (finite) Casimir force is exactly that obtained from $\sigma_{xx}$. The divergence of the Casimir energy density on the boundaries of the half-spaces does not affect the Casimir force because it is not caused by the cavity between them. A single planar slab of dielectric suffers from the same divergence of its Casimir surface energy, a phenomenon noted by Barton, using a completely different theoretical framework, for the case of a plasma sheet~\cite{bar05a,bar05b}. Although the Casimir forces on planar homogeneous dielectrics come out as finite, the divergences in the energy-momentum tensor cannot be dismissed as unobservable since the total energy-momentum tensor has a local significance as the source of the gravitational field. There is therefore still an issue of divergences to be explained here, though luckily it does not interfere with the prediction of a finite Casimir force.

Let us turn to the case of general inhomogeneous media, where $\varepsilon$ and $\mu$ are allowed to vary continuously in space. In these circumstances the Green tensor no longer has a clear decomposition into a bulk part and a scattered part. The regularization procedure advocated in standard Lifshitz theory~\cite{dzy61,LL} is the following:  before using the Green tensor to calculate the stress tensor at $\mathbf{r}$, subtract the bulk Green tensor for a \textit{homogeneous} medium whose constant values of  $\varepsilon$ and $\mu$ are equal to $\varepsilon(\mathbf{r})$ and $\mu (\mathbf{r})$. This recipe is designed to remove the contribution to the vacuum stress that does \textit{not} arise from material inhomogeneity. To our knowledge, this regularization has never been implemented in a solved example, and in using it for the solution presented in the next section we will find that it gives an infinite Casimir stress in the inhomogeneous dielectric. The cause of the difficulty may be that, while this regularization excludes the contribution to the stress from each infinitesimal (and therefore homogeneous) volume element of the dielectric, it includes the contribution made by the inhomogeneity down to arbitrarily small scales where the concept of a dielectric permittivity is invalid. Realistic models of $\varepsilon(\mathbf{r},i \xi)$ and $\mu(\mathbf{r},i \xi)$ should not vary significantly over distances comparable to the molecular size, but if continuous functions of $\mathbf{r}$ are chosen there will be some variation of $\varepsilon(\mathbf{r},i \xi)$ and $\mu(\mathbf{r},i \xi)$ over arbitrarily small distances, and this small-scale inhomogeneity will contribute to the Casimir stress if the standard regularization is used. One might therefore seek a regularization that removes the contribution to the stress from the small-scale inhomogeneity. Such a regularization can be performed by constructing a tensor $\mathbf{G}_0(\mathbf{r},\mathbf{r'},i \xi)$ that takes some accurate account of the small-scale inhomogeneity and reduces to the bulk Green tensor for a homogeneous medium when $\mathbf{r'}\rightarrow\mathbf{r}$; subtracting this from the exact Green tensor will exclude the small-scale inhomogeneity, hopefully yielding a finite stress tensor. 

To construct the tensor $\mathbf{G}_0(\mathbf{r},\mathbf{r'},i \xi)$ described above, we specialize to our case of interest, the one-dimensional inhomogeneity of the last section. Here we can work with the scalar electric and magnetic Green functions $\widetilde{g}_{E}(x,x')$ and $\widetilde{g}_{M}(x,x')$, and we require functions $\widetilde{g}_{0E}(x,x')$ and $\widetilde{g}_{0M}(x,x')$ that take accurate account of the small-scale inhomogeneity and reduce to (\ref{ghom}) when $x'\rightarrow x$. The medium at small scales is probed by waves with large transverse wave-vector components $u$ and $v$, and we wish to solve accurately (\ref{gEsim})--(\ref{gMsim}) when the waves reach the scale where the wavelength is much smaller than the distance over which $\varepsilon(x,i \xi)$ and $\mu(x,i \xi)$ are varying significantly. Looking at (\ref{gEsim})--(\ref{gMsim}) we see that these equations describe the regime in question when the following are satisfied:
\begin{gather}  
\left|\frac{1}{\varepsilon(x,i \xi)}\frac{d\varepsilon(x,i \xi)}{dx}\right|\ll w^2, \qquad  \left|\frac{1}{\mu(x,i \xi)}\frac{d\mu(x,i \xi)}{dx}\right|\ll w^2,   \label{approx}  \\[5pt]
w:=\sqrt{u^2+v^2+\varepsilon(x,i \xi)\mu(x,i \xi)\frac{\xi^2}{c^2}}.  \label{wdef2}
\end{gather}
Note $\varepsilon(x,i \xi)$ and $\mu(x,i \xi)$ are real for positive imaginary frequencies and for dissipative media they decrease monotonically to 1 as $\xi$ ranges from $0$ to $\infty$~\cite{LLcm}; the condition (\ref{approx}) is therefore unaffected by the $\xi$-dependence. Consider (\ref{gEsim}) without the delta function, the solution of which we denote by $g$:
\begin{equation}
\left(\partial_x\frac{1}{\mu(x,i \xi)}\partial_x-\frac{u^2+v^2}{\mu(x,i \xi)}-\frac{\xi^2}{c^2} \varepsilon(x,i \xi)\right)g =0, \label{g}
\end{equation}
An approximate solution of (\ref{g}) for slowly varying $\varepsilon(x,i \xi)$ and $\mu(x,i \xi)$ can be obtained by the WKB method~\cite{merz}; combined with (\ref{approx}) this yields the approximation
\begin{equation} \label{gsol}
g\approx C \sqrt{\frac{\mu(x,i \xi)}{w}}\, \exp\left(\pm\int^xw\,dx\right).
\end{equation}
Comparing (\ref{gsol}) with the homogeneous electric Green function (\ref{ghom}) and the reciprocity property (\ref{recipE}), we immediately obtain the desired $\widetilde{g}_{0E}(x,x')$. The magnetic function $\widetilde{g}_{0M}(x,x')$ follows similarly and we have~\cite{leo}
\begin{gather}
\widetilde{g}_{0E}(x,x')=-\frac{\sqrt{\mu(x,i \xi)\mu(x',i \xi)}}{2\sqrt{ww'}}\, \exp\left(-\left|\int^x_{x'}w\,dx\right|\right),   \label{g0E} \\[5pt]
 \widetilde{g}_{0M}(x,x')=-\frac{\sqrt{\varepsilon(x,i \xi)\varepsilon(x',i \xi)}}{2\sqrt{ww'}}\, \exp\left(-\left|\int^x_{x'}w\,dx\right|\right),   \label{g0M}  \\[5pt]
w':=\sqrt{u^2+v^2+\varepsilon(x',i \xi)\mu(x',i \xi)\frac{\xi^2}{c^2}}.  \label{w'def}
\end{gather}
The functions (\ref{g0E})--(\ref{g0M}) are a modification of the homogeneous Green functions (\ref{ghom}) that take account of the inhomogeneity of the medium over small distances where $\varepsilon(x,i \xi)$ and $\mu(x,i \xi)$ do not vary significantly. This small-scale inhomogeneity is experienced only by waves with very large $u$ and $v$. We can remove the contribution of this inhomogeneity to the Casimir stress by subtracting (\ref{g0E}) and (\ref{g0M}) from the Green functions $\widetilde{g}_{E}$ and $\widetilde{g}_{M}$; this gives the regularized Green functions
\begin{equation}  \label{greg}
\widetilde{g}_{\text{reg}E}:=\widetilde{g}_{E}-\widetilde{g}_{0E},  \qquad\widetilde{g}_{\text{reg}M}:=\widetilde{g}_{M}-\widetilde{g}_{0M}.
\end{equation}

It is clear from the foregoing that the small-scale inhomogeneity can also be removed from the Casimir stress by introducing a cutoff for large wave vectors. One might consider doing this by means of spatial dispersion, with $\varepsilon$ and $\mu$ going to $1$ for wavelengths on the molecular scale; however, wave-vector components can only be introduced along homogeneous directions, as we have been able to do in the $yz$-plane by means of (\ref{four}). A cutoff in the transverse wave vector components $u$ and $v$ will yield a finite Casimir force, but the status of such cutoff-dependent results is disputed (see, for example, the previously cited literature on the spherical Casimir effect).

The Casimir force (\ref{F}) depends only on the component $\sigma_{xx}$ of the stress tensor, because of the translation symmetry in the $yz$-plane. In terms of the scalar Green functions, the stress tensor is computed from (\ref{stress})--(\ref{BB}), (\ref{four}), and (\ref{Gsol}), with the Green functions in (\ref{Gsol}) regularized either by (\ref{greg}) or according to the standard recipe based on (\ref{ghom}). Inversion of the Fourier transform (\ref{four}) leads to a double integral $\int_{-\infty}^\infty du\int_{-\infty}^\infty dv$, but the integrand for $\sigma_{xx}$ only contains $u$ and $v$ in the combination $u^2+v^2$, so we can set $\int_{-\infty}^\infty du\int_{-\infty}^\infty dv\rightarrow 2\pi\int_{0}^\infty du$ and $u^2+v^2\rightarrow u^2$. The result for $\sigma_{xx}$ is
\begin{gather}
\sigma_{xx}(x)=-\frac{\hbar}{4\pi^2}\int_{0}^\infty du\int_{0}^\infty d\xi\, \widetilde{\sigma}_{xx}(x,u,i\xi),  \label{xx1}  \\[5pt]
\widetilde{\sigma}_{xx} (x,u,i\xi):=u\left[w^2\left(\frac{\widetilde{g}_{\text{reg}E}}{\mu}+\frac{\widetilde{g}_{\text{reg}M}}{\varepsilon}\right)-\frac{1}{\mu}\partial_x\partial_{x'}\widetilde{g}_{\text{reg}E}-\frac{1}{\varepsilon}\partial_x\partial_{x'}\widetilde{g}_{\text{reg}M}\right]_{\stackrel{\scriptstyle v=0}{x'=x}}.    \label{xx2}
\end{gather}
We obtain from (\ref{F}) the ($x$-directed) Casimir force per unit area $F_A$ on a slab of material bounded by $x=x_1$ and $x=x_2$ ($x_2>x_1$):
\begin{equation}  \label{FA}
F_A=\sigma_{xx}(x_2)-\sigma_{xx}(x_1),
\end{equation}
and the ($x$-directed) force per unit volume (\ref{f}) simplifies to
\begin{equation}   \label{f2}
f=\frac{d\sigma_{xx}}{dx}.
\end{equation}

In the following section we calculate the Casimir stress (\ref{xx1})--(\ref{xx2}) for a simple model. As well as employing the  the standard regularization~\cite{LL} based on (\ref{ghom}), we also show the result when the regularization (\ref{greg}) is used.

\section{Example}  \label{example}
The formalism of the last section was set up for dispersive media, with  $\varepsilon(x,i \xi)$ and $\mu(x,i \xi)$ taken as functions of (imaginary) frequency $\xi$. As well as being essential for an accurate calculation of the Casimir force for specific materials, the inclusion of the measured dispersion properties provides a physical cutoff in the imaginary-frequency integral in (\ref{xx1}). We will be encountering divergence problems with the formula (\ref{xx1}), but none of these divergences can be cured by the introduction of (temporal) dispersion since they always involve a failure of the integration over the transverse wave vector $u$ to converge. What we require is a simple model to test the theory of the Casimir effect for inhomogeneous media; since dispersion does not resolve the theoretical issues we do not include it in our results. In addition, we assume the medium is a dielectric with no magnetic response ($\mu=1$), so that the inhomogeneous medium is completely described by a permittivity function $\varepsilon(x)$. 

\begin{figure}[t]
\includegraphics[width=18.0pc]{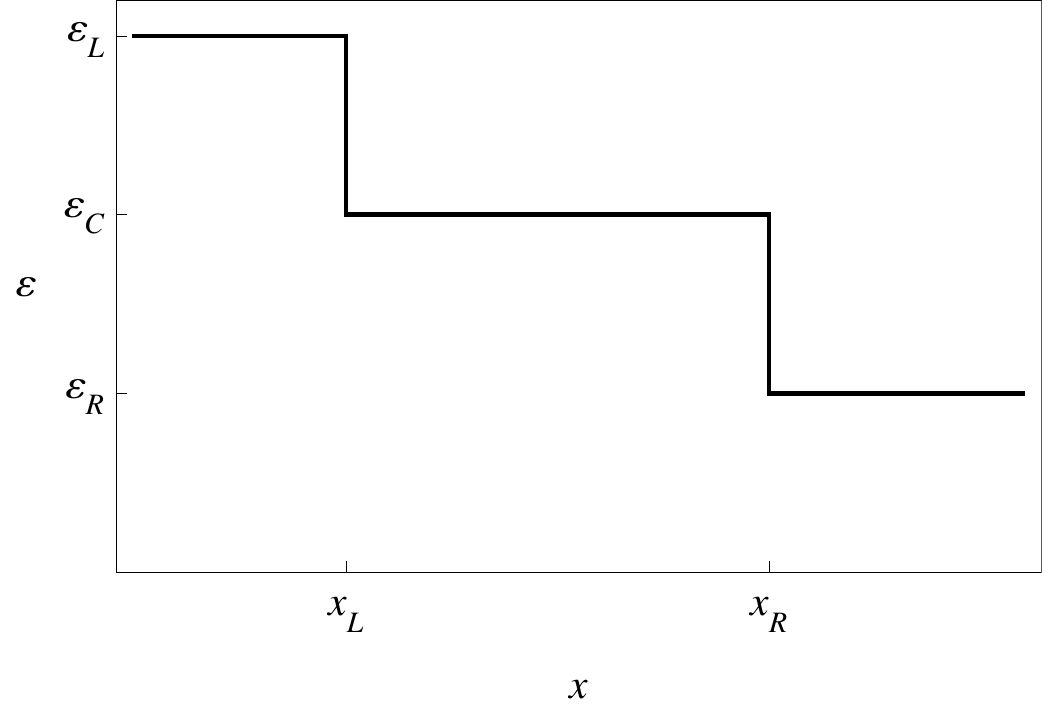}
\hspace{5mm}
\includegraphics[width=18.0pc]{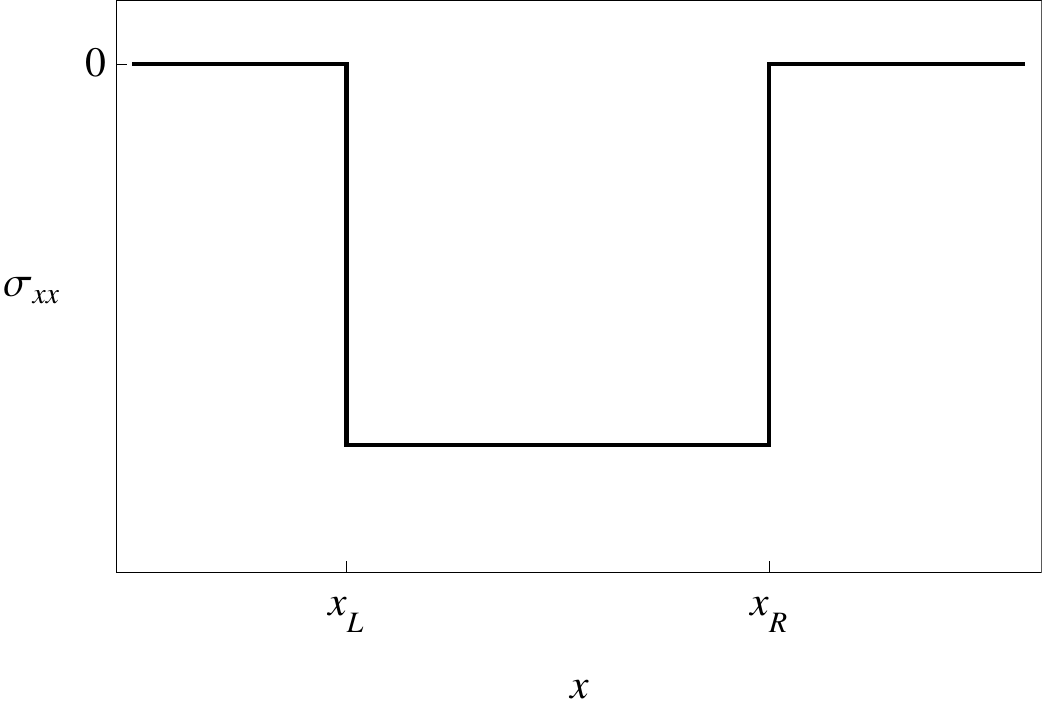}
\caption{The left plot shows the permittivity for a central homogeneous dielectric ($\varepsilon= \varepsilon_C$) filling $x_L<x<x_R$, flanked by two dielectric half-spaces with $\varepsilon= \varepsilon_L$ (left) and $\varepsilon= \varepsilon_R$ (right). The right plot shows the resulting Casimir stress $\sigma_{xx}$, zero in the half spaces and negative (constant) in between. The Casimir force repels the half-spaces from each other.}   \label{fig1}
\end{figure}

In choosing a function $\varepsilon(x)$, it is interesting to consider a generalization of an arrangement that gives a repulsive Casimir force. The classic configuration of Casimir theory is that of parallel half-spaces made of homogeneous dielectrics, separated by a third homogeneous dielectric~\cite{dzy61}. If the permittivities $\varepsilon_L$,  $\varepsilon_R$ and $\varepsilon_C$ of the left, right and central dielectrics obey
\begin{equation}  \label{con}
\varepsilon_L>\varepsilon_C>\varepsilon_R,
\end{equation}
then the Casimir force on the half-spaces is repulsive~\cite{dzy61}. (The left and right half-spaces can of course be interchanged to the same effect.) The arrangement (\ref{con}) of permittivities is depicted in Fig.~\ref{fig1}, together with the resulting Casimir stress $\sigma_{xx}$ that determines the force. The stress  $\sigma_{xx}$ is zero in the half-spaces and has a constant negative value between them. An  integration of the stress tensor over the boundary surface of each half space gives the Casimir force (\ref{F}) acting on it; the non-zero force component $F_x$ is clearly negative for the left half-space and positive for right half-space, so they are repelled from each other. The repulsive force per unit area on each half space is $|\sigma_{xx}|$. The result is the same if the outer dielectrics have a finite, macroscopic thickness, since in the force per unit area (\ref{FA}), $\sigma_{xx}$ will effectively be zero on the outer boundaries of the plates while having the same value between them as for infinite thickness. For real, dispersive dielectrics the permittivities will change with frequency and to obtain a repulsive force for a given plate separation, (\ref{con}) must hold over an appropriate range of imaginary frequencies $i\xi$~\cite{dzy61}. (If (\ref{con}) holds only for a finite range of $\xi$ then, depending on the separation, the force may be repulsive or attractive force when the integration over $\xi$ is performed~\cite{dzy61}). This remarkable prediction~\cite{dzy61} has now been verified experimentally by the measurement of a repulsive Casimir force between a gold sphere and a silica plate immersed in bromobenzene~\cite{mun09}.

\begin{figure}[t]
\begin{center}
\includegraphics[width=18.0pc]{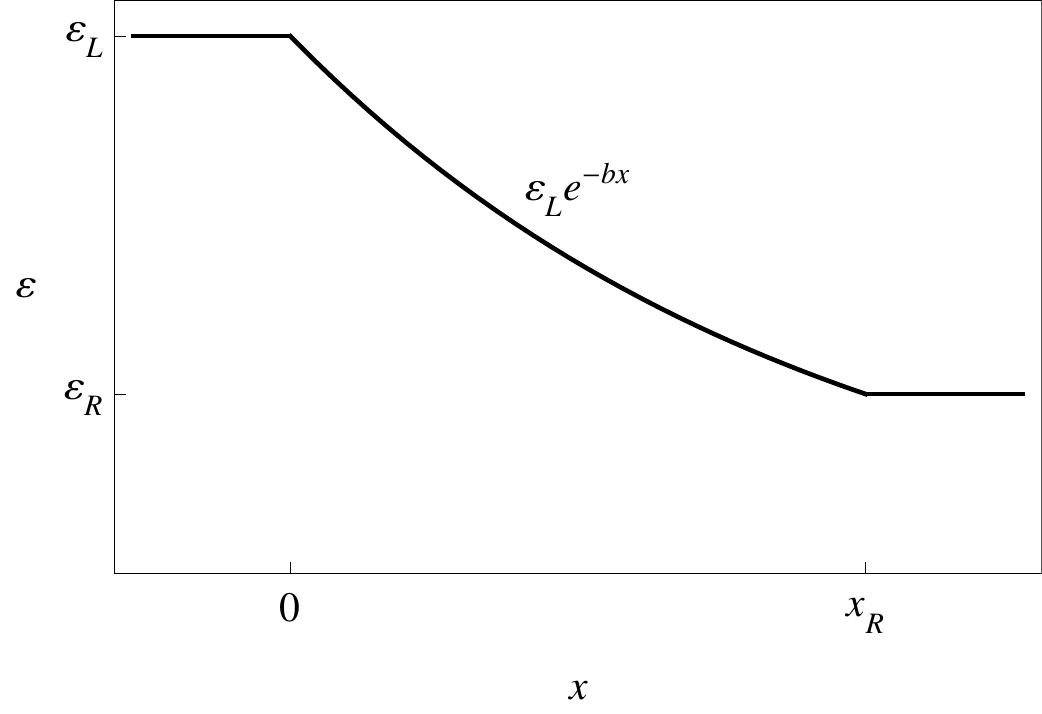}
\end{center}
\caption{A modification of the permittivity in Fig.~\ref{fig1} that changes the material in the central region so that $\varepsilon$ is continuous everywhere. With the choice of the displayed exponentially decreasing permittivity in the inhomogeneous region $0<x<x_R$, the integrand (\ref{xx2}) in the Casimir stress (\ref{xx1}) can be found analytically.}   \label{fig2}
\end{figure}

Because of the intrinsic interest of the configuration shown in Fig.~\ref{fig1}, we consider a generalization of it where the permittivity is a continuous function of $x$ so that the central dielectric is inhomogeneous---see Fig.~\ref{fig2}. Our choice of the monotonically decreasing $\varepsilon(x)$ in the central region was determined by requiring that the equations (\ref{gEsim})--(\ref{gMsim}) for the Green functions be solvable analytically. We obtain analytic solutions for the Green functions with the permittivity
\begin{equation}  \label{perm}
\varepsilon(x)=\varepsilon_Le^{-bx}, \qquad \text{$\varepsilon_L,b$ positive constants,}
\end{equation}
and we join this continuously to homogeneous half-spaces on the left at $x=0$ and on the right at $x=x_R$, as shown in Fig.~\ref{fig2}. This example provides an interesting test of intuition about the Casimir effect, since from the result in Fig.~\ref{fig1} one might guess that the parts of the inhomogeneous medium close to $x=0$ and $x=x_R$ experience a Casimir force that repels them from each other. This would imply that $\sigma_{xx}$ is u-shaped in the region $0<x<x_R$ so that the force per unit volume (\ref{f2}) changes from negative to positive as one moves from $x=0$ to $x=x_R$.  One could imagine a solid dielectric with the permittivity of Fig.~\ref{fig2} being produced through a manufacturing process that allows the density of the solid to be varied continuously as it is constructed; the variation of the permittivity in Fig.~\ref{fig2} would thus be caused by a similar variation in the solid's density. 

We proceed to find the scalar Green functions with the permittivity (\ref{perm}) and $\mu=1$. As seen from (\ref{xx2}), we can set $v=0$, whereupon (\ref{gEsim}) and (\ref{gMsim}) for the permittivity (\ref{perm}) are
\begin{gather}
\frac{d^2\widetilde{g}_E}{dx^2}-\left(u^2+\varepsilon_Le^{-bx}\frac{\xi^2}{c^2}\right)\widetilde{g}_E =\delta(x-x'), \label{gEex} \\[5pt]
\frac{d}{dx}\frac{1}{\varepsilon_L}e^{bx}\frac{d}{dx}\widetilde{g}_M-\left(\frac{u^2}{\varepsilon_L}e^{bx}+\frac{\xi^2}{c^2}\right)\widetilde{g}_M =\delta(x-x'). \label{gMex}
\end{gather}
Both (\ref{gEex}) and (\ref{gMex}) can be recast as the modified Bessel equation with an added delta-function term. Taking first the electric Green function (\ref{gEex}), we make the variable change
\begin{equation}  \label{sdef}
s(x):=\frac{2\xi\sqrt{\varepsilon_L}}{bc}e^{-bx/2} \quad \Longrightarrow \qquad x(s)=-\frac{2}{b}\ln\left(\frac{bcs}{2\xi\sqrt{\varepsilon_L}}\right).
\end{equation}
The delta function $\delta(x-x')$ in (\ref{gEex}) is related to $\delta(s-s')$ by
\begin{equation}  \label{dtrans}
\delta(x(s)-x'(s'))=\frac{1}{\left|\frac{d}{ds}\left[x(s)-x'(s')\right]\right|_{s=s'}}\delta(s-s')=\frac{1}{2}bs'\delta(s-s'),
\end{equation}
and the other terms in (\ref{gEex}) are easily expressed in terms of $s$, yielding
\begin{equation} \label{gEexsim}
s^2\frac{d^2\widetilde{g}_E}{ds^2}+s\frac{d\widetilde{g}_E}{ds}-\left(s^2+\frac{4u^2}{b^2}\right)\widetilde{g}_E =\frac{2s'}{b}\delta(s-s').
\end{equation}
With the delta-function term removed, (\ref{gEexsim}) is the modified Bessel equation, independent solutions for which are provided by the modified Bessel functions $I_{2u/b}(s)$ and $K_{2u/b}(s)$~\cite{abr}. The general solution of (\ref{gEexsim}) can be written
\begin{equation}  \label{gEsol}
\widetilde{g}_E=c_{E1}I_{2u/b}(s)+c_{E2}K_{2u/b}(s)+\frac{2}{b}\left[I_{2u/b}(s)K_{2u/b}(s')-I_{2u/b}(s')K_{2u/b}(s)\right]\theta(s-s').
\end{equation}
The term in (\ref{gEsol}) proportional to the Heavyside theta function $\theta(s-s')$ generates the delta function on the right-hand side of (\ref{gEexsim}), as is verified by substitution and use of the identities
\begin{gather}
\frac{d}{ds}I_{n}(s)=\frac{1}{2}\left[I_{n-1}(s)+I_{n+1}(s)\right], \qquad \frac{d}{ds}K_{n}(s)=-\frac{1}{2}\left[K_{n-1}(s)+K_{n+1}(s)\right],  \\[5pt]
\frac{1}{2}sI_{n+1}(s)+nI_{n}(s)-\frac{1}{2}sI_{n-1}(s)=0, \\[5pt]
\frac{1}{2}sK_{n+1}(s)-nK_{n}(s)-\frac{1}{2}sK_{n-1}(s)=0.
\end{gather}
It is important to note that the case $s>s'$, for which the term in (\ref{gEsol}) proportional to $\theta(s-s')$ is non-zero, corresponds to $x<x'$, as is seen from the definition (\ref{sdef}); similarly, $s<s'$ implies $x>x'$. After the same variable change (\ref{sdef}), the magnetic Green function equation (\ref{gMex}) can be written
\begin{equation} \label{gMexsim}
s^2\frac{d^2}{ds^2}\left(\frac{1}{s}\widetilde{g}_M\right)+s\frac{d}{ds}\left(\frac{1}{s}\widetilde{g}_M\right)-\left(s^2+1+\frac{4u^2}{b^2}\right)\left(\frac{1}{s}\widetilde{g}_M\right)=\frac{bc^2s'^2}{2\xi^2}\delta(s-s').
\end{equation}
Without the delta-function term, (\ref{gMexsim}) is the modified Bessel equation for the function $\widetilde{g}_M/s$. The general solution of (\ref{gMexsim}) can immediately be written down by comparing it with (\ref{gEexsim}) and the latter's general solution (\ref{gEsol}):
\begin{gather}  
\widetilde{g}_M=c_{M1}sI_{\nu}(s)+c_{M2}sK_{\nu}(s)+\frac{bc^2ss'}{2\xi^2}\left[I_{\nu}(s)K_{\nu}(s')-I_{\nu}(s')K_{\nu}(s)\right]\theta(s-s'),  \label{gMsol}  \\[5pt]
\nu:=\sqrt{1+\frac{4u^2}{b^2}}.
\end{gather}

Equations (\ref{gEsol}) and (\ref{gMsol}) are the solutions for the Green functions in the inhomogeneous region in Fig.~\ref{fig2}. The solutions in the homogeneous regions $x<0$ and $x>x_R$ are given by (\ref{ghomtot}), once we take account that the waves from the dipole at $x'$ are outgoing to $x=\pm\infty$; this means that for $x<0$ the coefficients of $e^{-wx}$ in the Green functions (\ref{ghomtot}) vanish, whereas for $x>x_R$ the coefficients of $e^{wx}$ vanish. There are thus four unknown constants in the complete solutions for both the electric and magnetic Green functions, and these constants are determined by imposing the continuity of (\ref{boundary}) at the interfaces $x=0$ and $x=x_R$. This is a mechanical exercise that yields very lengthy and unenlightening expressions for the constants. The resulting exact Green functions are then regularized, either by (\ref{greg}) or by the standard recipe based on (\ref{ghom}), and the integrand (\ref{xx2}) in the Casimir stress (\ref{xx1}) is found. The double integral in (\ref{xx1}) must be evaluated numerically and we proceed to exhibit the results for specific values of the parameters in Fig.~\ref{fig2}.

\begin{figure}[t]
\begin{center}
\includegraphics[width=18.0pc]{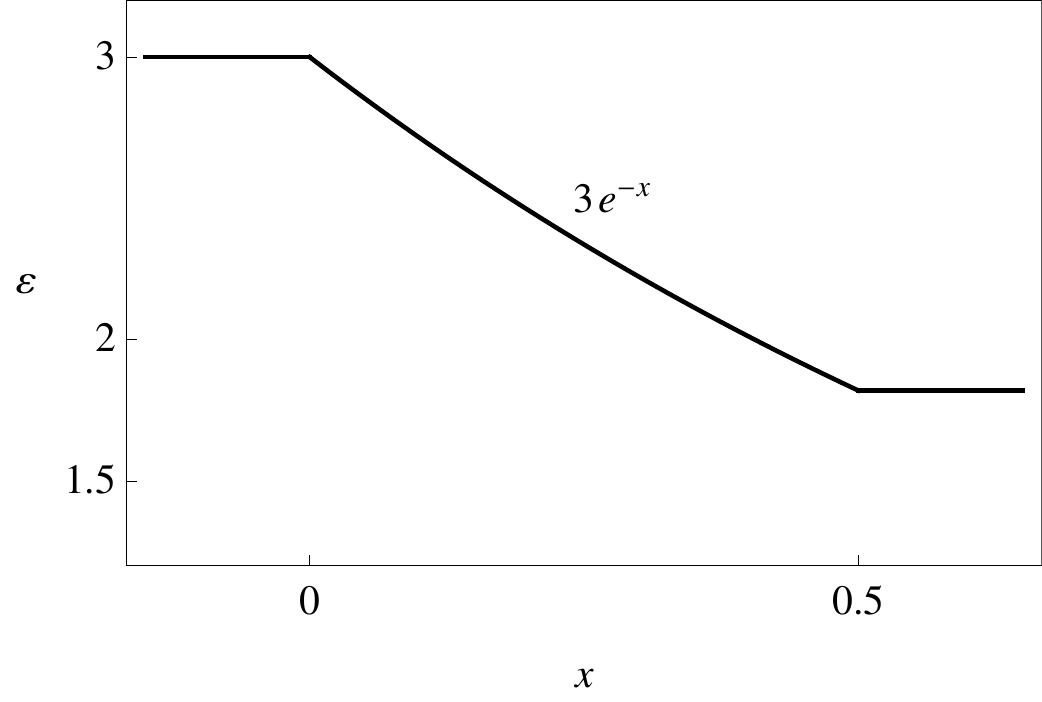}
\end{center}
\caption{The permittivity of Fig.~\ref{fig2} for the parameter values $\varepsilon_L=3$, $b=1$ and $x_R=0.5$. This gives $\varepsilon_R=1.82$.}   \label{fig3}
\end{figure}

\begin{figure}[t]
\includegraphics[width=18.0pc]{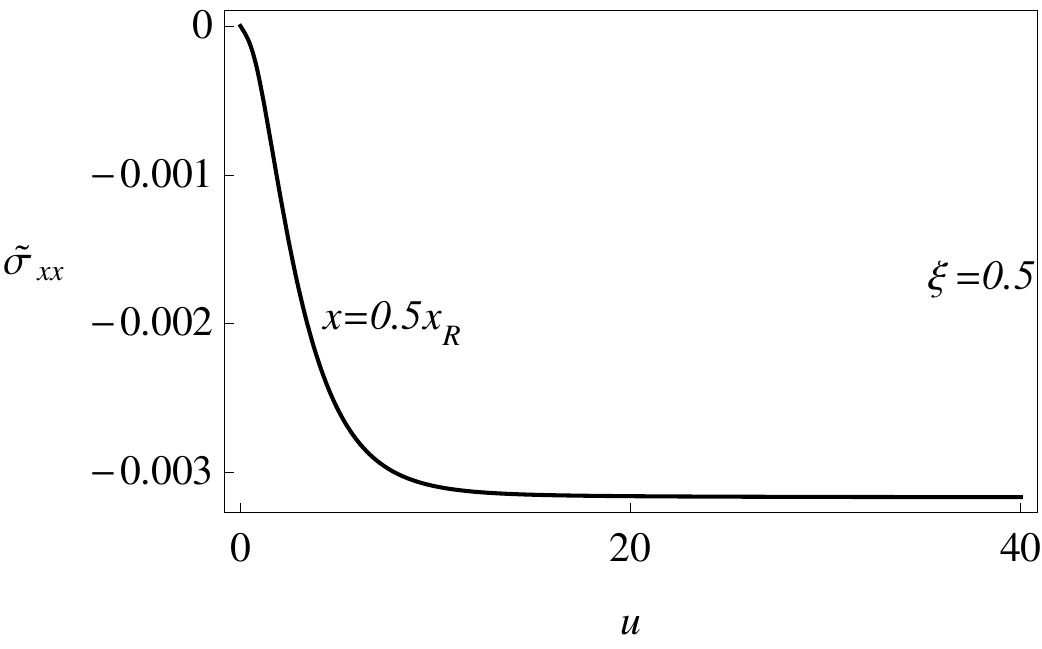}
\hspace{5mm}
\includegraphics[width=18.0pc]{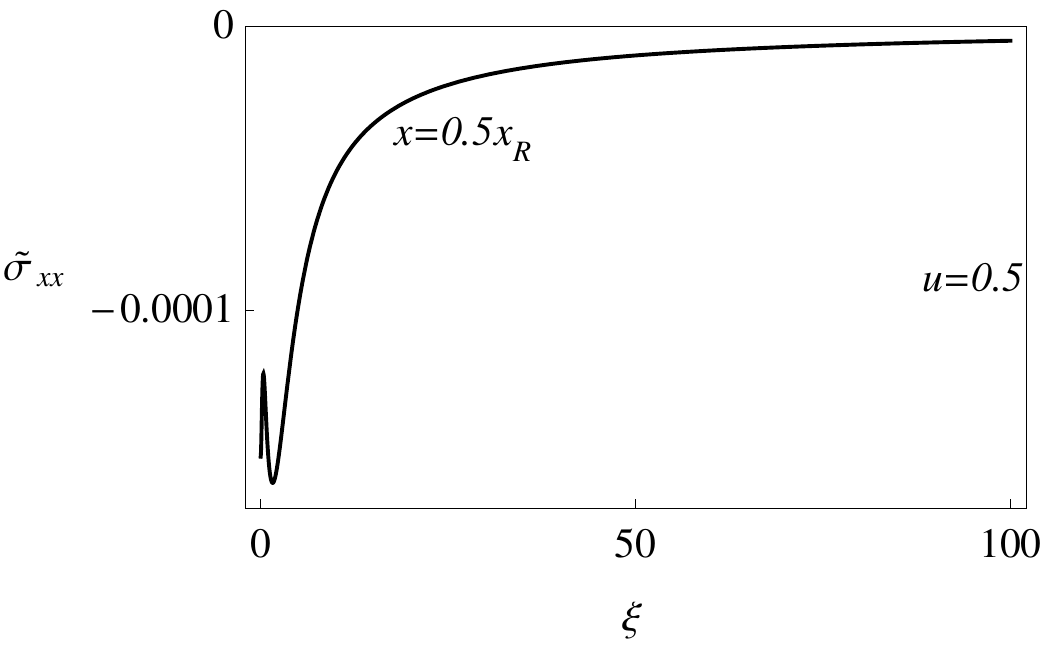}
\caption{The integrand (\ref{xx2}) in the Casimir stress (\ref{xx1}) with the parameters of Fig.~\ref{fig3} and $\hbar=c=1$, where the Green functions are regularized according to the standard prescription of Lifshitz theory~\cite{dzy61,LL}. The left figure plots $\widetilde{\sigma}_{xx}$ as a function of $u$ with $x=0.5x_R$ and $\xi=0.5$, while the right figure plots $\widetilde{\sigma}_{xx}$ as a function of $\xi$ for the same $x$, with $u=0.5$. The $\xi$-plot goes to zero as $\xi\rightarrow\infty$ but the $u$-plot shows a decrease to a constant negative value as $u\rightarrow\infty$, giving an infinite Casimir stress $\sigma_{xx}$. Similar plots for other $x$ values inside the inhomogeneous region $0<x<x_R$ show the same qualitative behaviour; the standard regularization therefore gives an infinite Casimir stress $\sigma_{xx}$ throughout the inhomogeneous medium.}   \label{fig4}
\end{figure}

\begin{figure}[t]
\includegraphics[width=18.0pc]{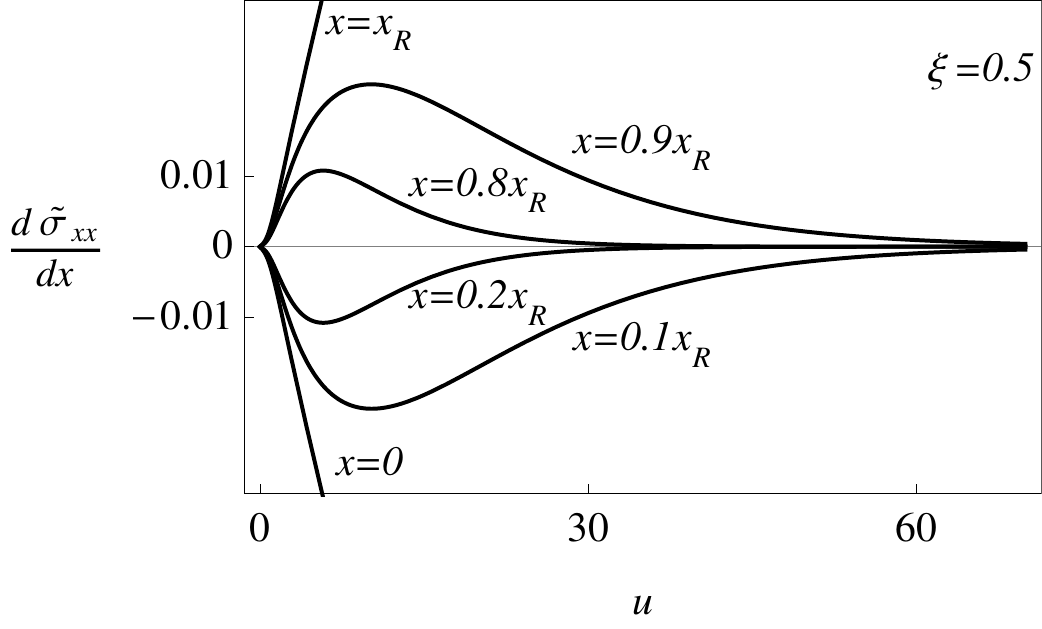}
\hspace{5mm}
\includegraphics[width=18.0pc]{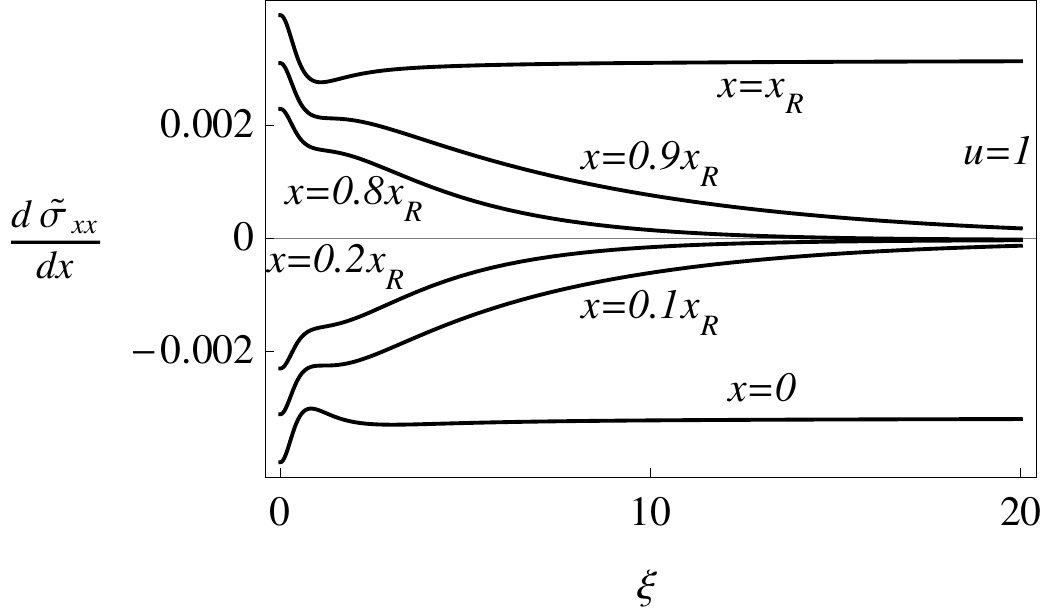}
\caption{The derivative of the integrand (\ref{xx2}) with the parameters of Fig.~\ref{fig3} and $\hbar=c=1$, where the Green functions are regularized according to the standard prescription of Lifshitz theory~\cite{dzy61,LL}. The left figure plots $d\widetilde{\sigma}_{xx}/dx$ as a function of $u$ for four values of $x$, with $\xi=0.5$, while the right figure plots $d\widetilde{\sigma}_{xx}/dx$ as a function of $\xi$ for the same $x$ values, with $u=0.5$. It is clear that the plots for $x=0$ and $x=x_R$ fail to approach zero for large $u$ or for large $\xi$, so integration of $d\widetilde{\sigma}_{xx}/dx$ with respect to $u$ and $\xi$ yields diverging values for the force per unit volume $f$ at these two positions. The values of $x$ chosen in the plots are close to or on the interfaces $x=0$ and $x=x_R$; similar plots for $x$ near the center of the inhomogeneous region give curves with pronounced oscillations, as shown in Fig.~\ref{fig6}.}   \label{fig5}
\end{figure}

\begin{figure}[t]
\includegraphics[width=18.0pc]{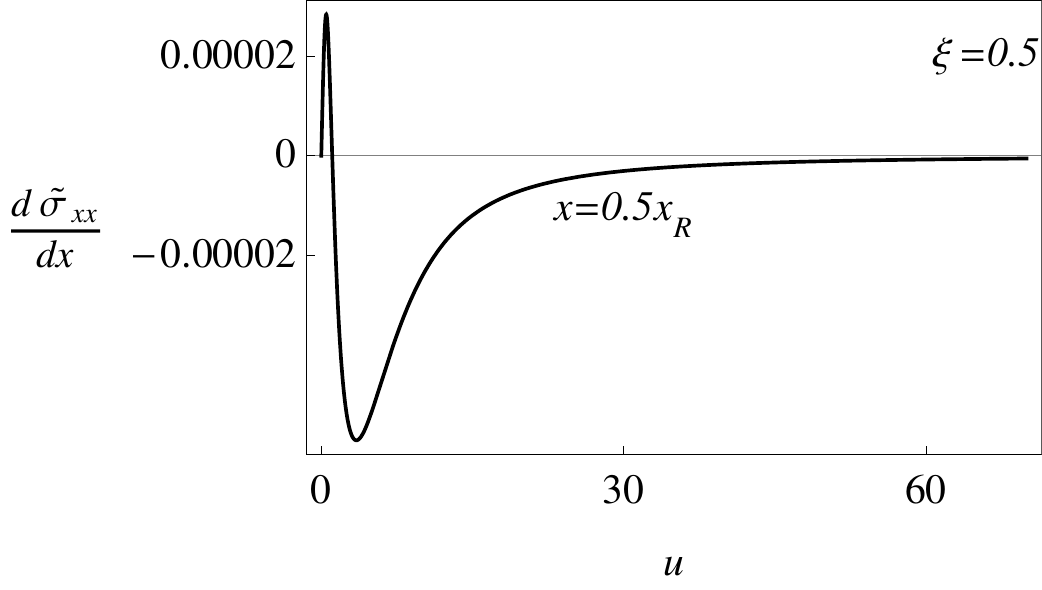}
\hspace{5mm}
\includegraphics[width=18.0pc]{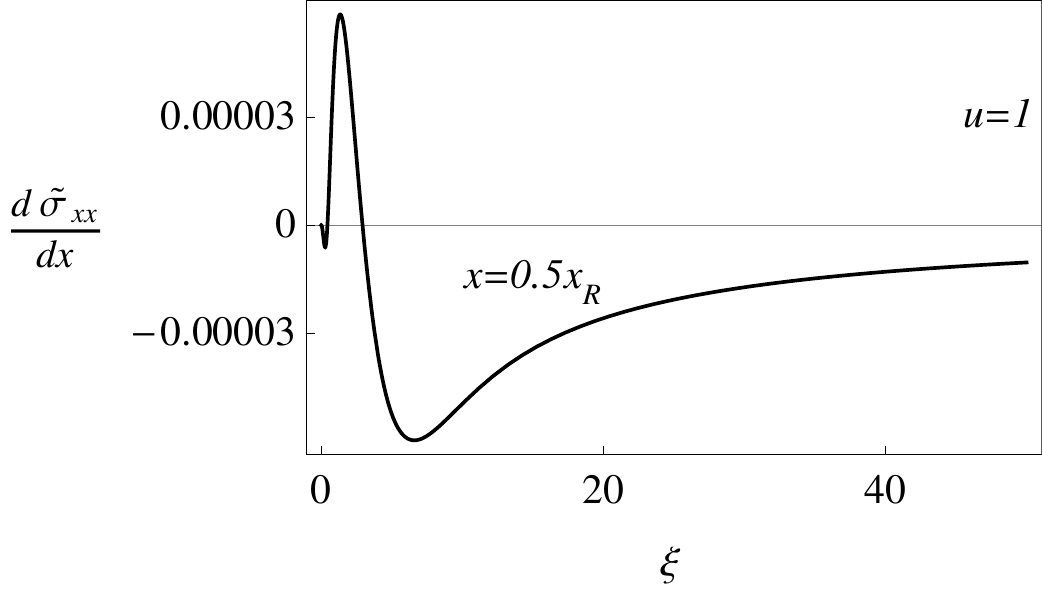}
\caption{The same as Fig.~\ref{fig5} but for the single position $x=0.5x_R$ at the centre of the inhomogeneous region. The curves are more oscillatory that those in Fig.~\ref{fig5} for positions close to or on the interfaces $x=0$ and $x=x_R$.}   \label{fig6}
\end{figure}

The first set of parameteter values is depicted in Fig.~\ref{fig3}, which shows the permittivity distribution of Fig.~\ref{fig2} with $\varepsilon_L=3$, $b=1$ and $x_R=0.5$, so that $\varepsilon_R=1.82$. We show first the result of using the regularization prescription for inhomogeneous media recommended in standard Lifshitz theory~\cite{dzy61,LL}. Recall that the standard regularization is to subtract from the total Green functions, at each $x$, the Green functions for a homogeneous medium whose constant value of $\varepsilon$ is equal to $\varepsilon(x)$. Figure~\ref{fig4} shows the behaviour of the integrand $\widetilde{\sigma}_{xx}$ if this regularization is used. The figure plots the resulting $\widetilde{\sigma}_{xx}$ as a function of $u$ with $\xi$ fixed, and as a function of $\xi$ with $u$ fixed, for $x=0.5x_R$. The $u$-plot for fixed $\xi$ approaches a constant negative value as $u\rightarrow\infty$,  giving an infinite value for the integral (\ref{xx1}) at this value of $x$, and this is the behaviour for any  $x$ inside the inhomogeneous medium $0<x<x_R$. The standard regularization thus gives an infinite Casimir stress $\sigma_{xx}$ throughout the inhomogeneous region. 

One might consider falling back on the fact that the Casimir force (\ref{FA}) on a slab of the dielectric is given by subtracting the stress at the two boundaries of the slab, and that a subtraction of two infinite values of the stress may give a finite answer. But one is free to consider a slab with one boundary in the inhomogeneous region of Fig.~\ref{fig3} and the other boundary in one of the homogeneous regions. With the standard regularization the force (\ref{FA}) on any such slab is infinite, since $\sigma_{xx}$ in the homogeneous regions is zero. Moreover, the legitimacy of an additional regularization to remove the infinite stress visible in Fig.~\ref{fig4} is questionable, since according to the standard theory~\cite{dzy61,LL} there should be no such infinity. As discussed in some detail in the last section, the expedient of additional regularizations on a case by case basis is a matter of continuing debate and disagreement. Nevertheless, for information purposes we consider the maneuver of taking the derivative in (\ref{f2}) inside the integrations in the infinite $\sigma_{xx}$; we can then differentiate the finite integrand $\widetilde{\sigma}_{xx}$ and afterwards perform the integrations to obtain a value for the force per unit area $f$. This could remove an $x$-independent infinity in $\sigma_{xx}$ and give a finite expression for $f$. Figures~\ref{fig5} and~\ref{fig6} show the behaviour of $d\widetilde{\sigma}_{xx}/dx$ at different values of $x$. It is clear from Fig.~\ref{fig5} that  $d\widetilde{\sigma}_{xx}/dx$ at the interfaces $x=0$ and $x=x_R=0.5$ is such that its integration would give a diverging value for the force per unit area $f$ at these positions. The plots of $d\widetilde{\sigma}_{xx}/dx$ inside the inhomogeneous region $0<x<x_R$ look more promising in that they approach zero for $u\rightarrow\infty$ and  $\xi\rightarrow\infty$. However, we find that the curves for $d\widetilde{\sigma}_{xx}/dx$ as a function of $\xi$ approach zero rather slowly as $\xi\rightarrow\infty$. Our attempts to numerically integrate $d\widetilde{\sigma}_{xx}/dx$ for a range of $x$ in $0<x<x_R$ produced values that varied significantly with small changes of $x$ and it is not clear that the integrals converge. In any case the obvious fact that the integration of $d\widetilde{\sigma}_{xx}/dx$ gives a diverging force per unit area on the interfaces  $x=0$ and $x=x_R$ shows that the standard regularization does not give an acceptable solution to the problem. 

\begin{figure}[t]
\includegraphics[width=18.0pc]{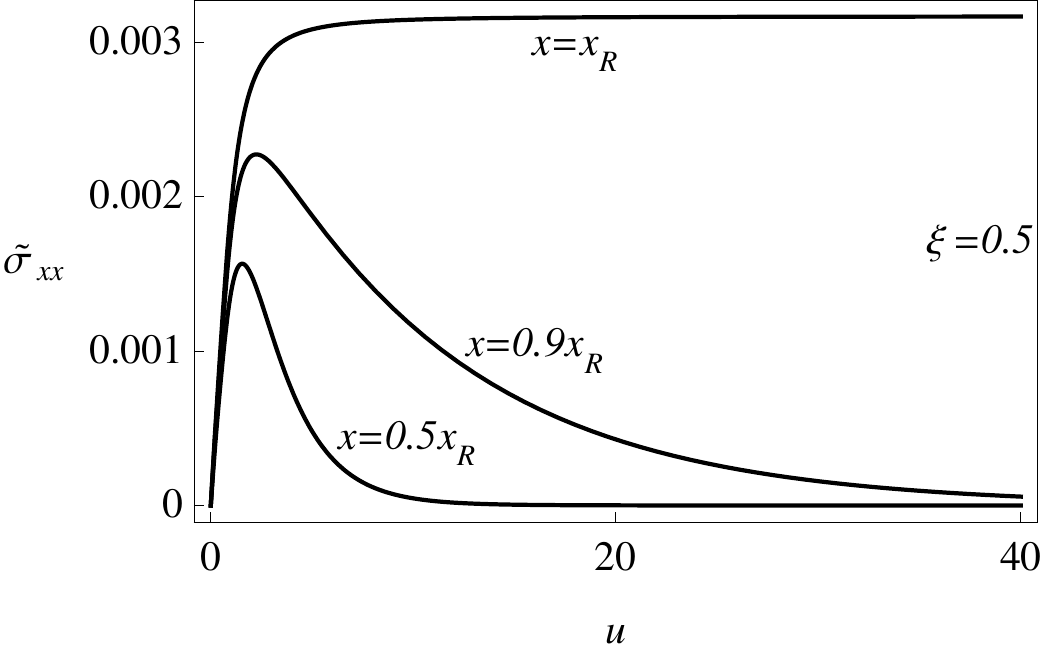}
\hspace{5mm}
\includegraphics[width=18.0pc]{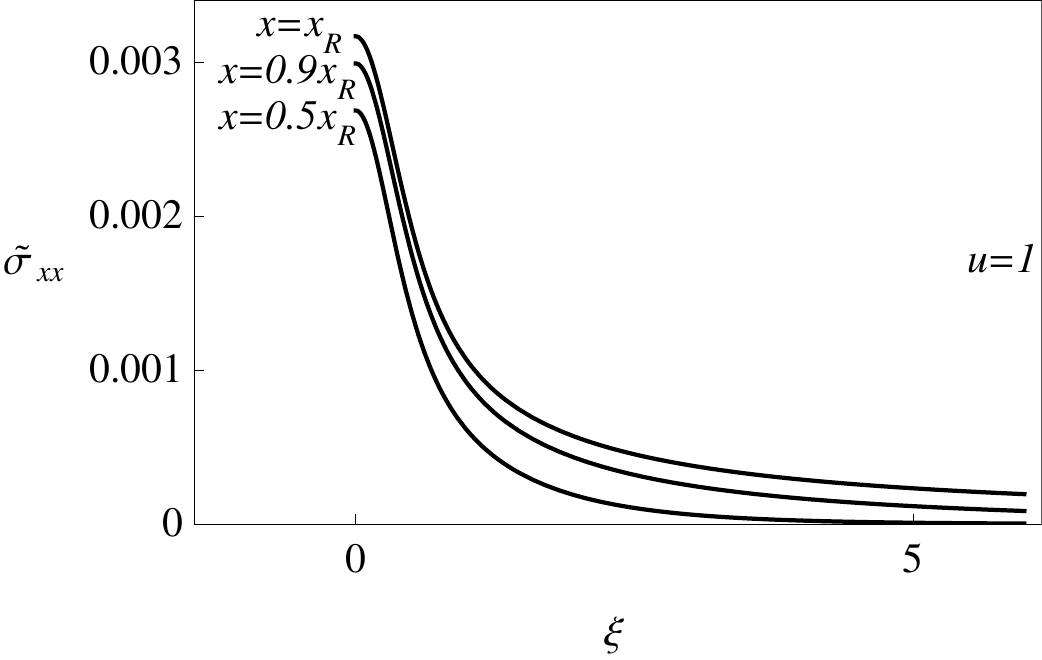}
\caption{The integrand (\ref{xx2}) in the Casimir stress (\ref{xx1}) with the parameters of Fig.~\ref{fig3} and $\hbar=c=1$, where the Green functions are regularized by (\ref{greg}) and (\ref{g0E})--(\ref{g0M}). The left figure plots $\widetilde{\sigma}_{xx}$ as a function of $u$ for three values of $x$, with $\xi=0.5$. For $x=x_R=0.5$ the integrand attains a constant positive value as $u\rightarrow\infty$, giving an infinite Casimir stress $\sigma_{xx}$ at this position. The right figure plots $\widetilde{\sigma}_{xx}$ as a function of $\xi$ for the same three values of $x$, with $u=1$; the curves approach zero as $\xi\rightarrow\infty$.}   \label{fig7}
\end{figure}

\begin{figure}[t]
\includegraphics[width=18.0pc]{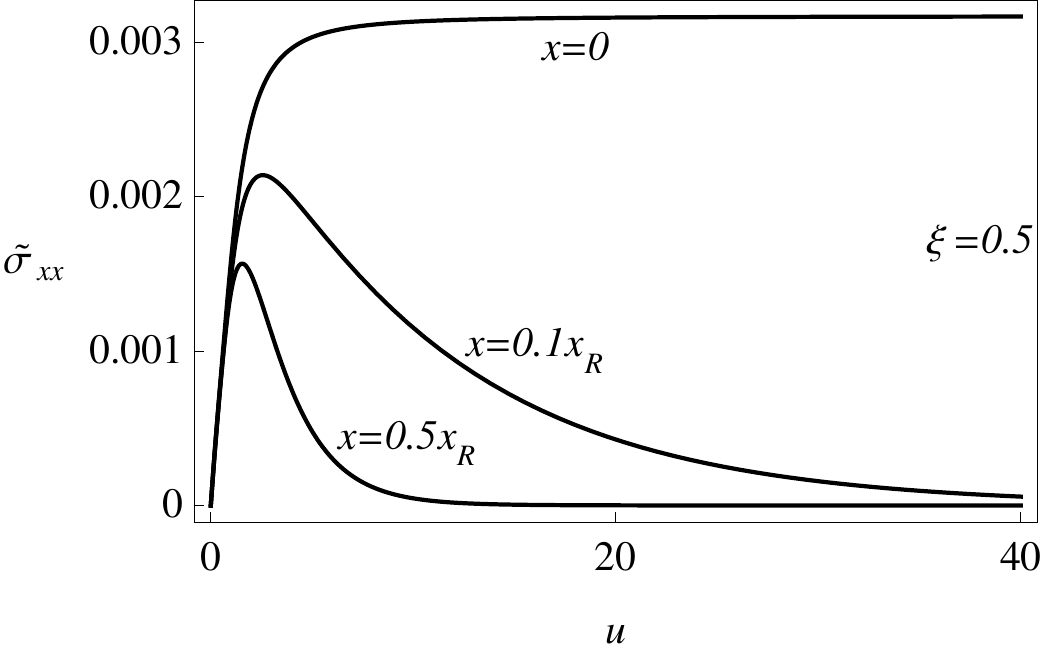}
\hspace{5mm}
\includegraphics[width=18.0pc]{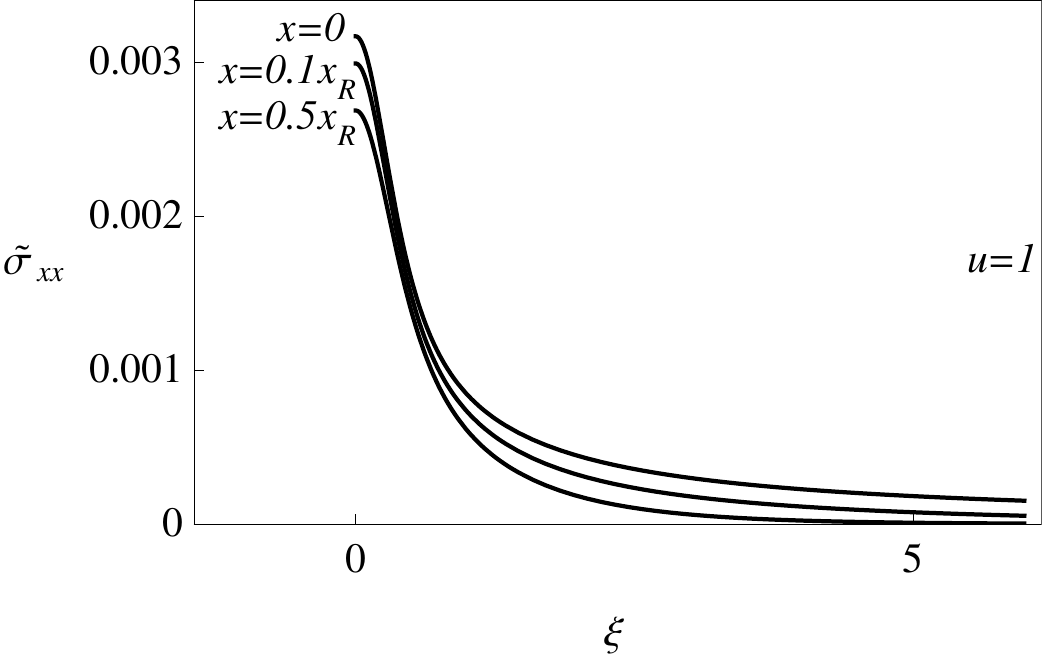}
\caption{The same as Fig.~\ref{fig7}, but with values of $x$ less than or equal to $0.5x_R=0.25$. For $x=0$ the behaviour of the $u$-plot is similar to that for $x=x_R$ in Fig.~\ref{fig7}, giving an infinite Casimir stress $\sigma_{xx}$ at this position.}   \label{fig8}
\end{figure}

\begin{figure}[t]
\includegraphics[width=18.0pc]{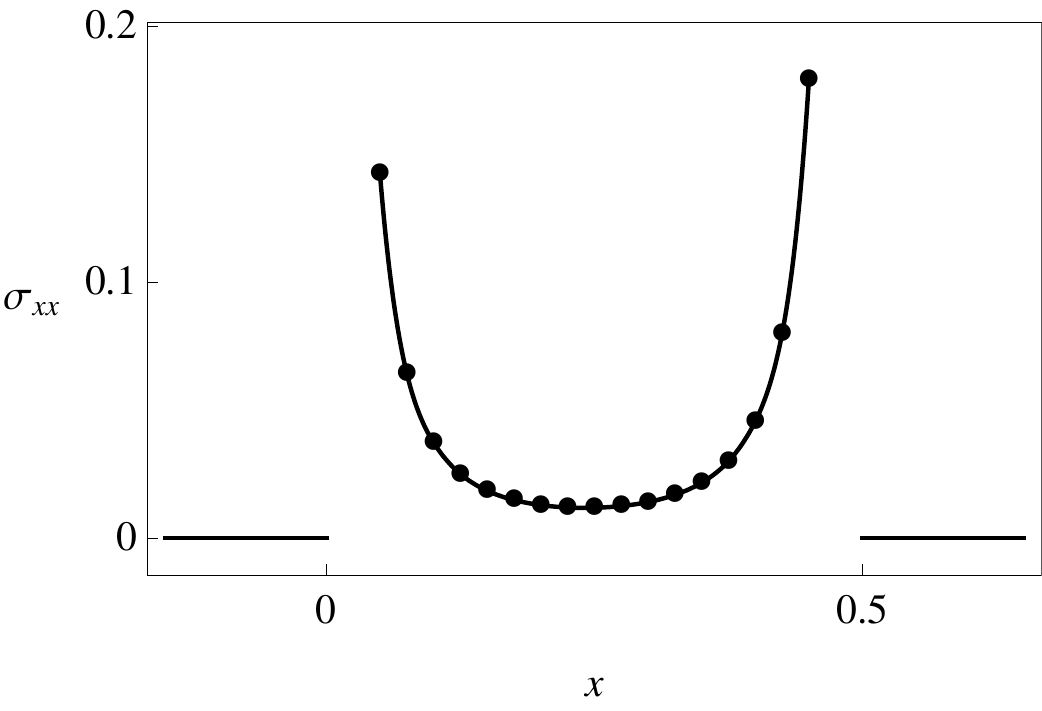}
\hspace{5mm}
\includegraphics[width=18.0pc]{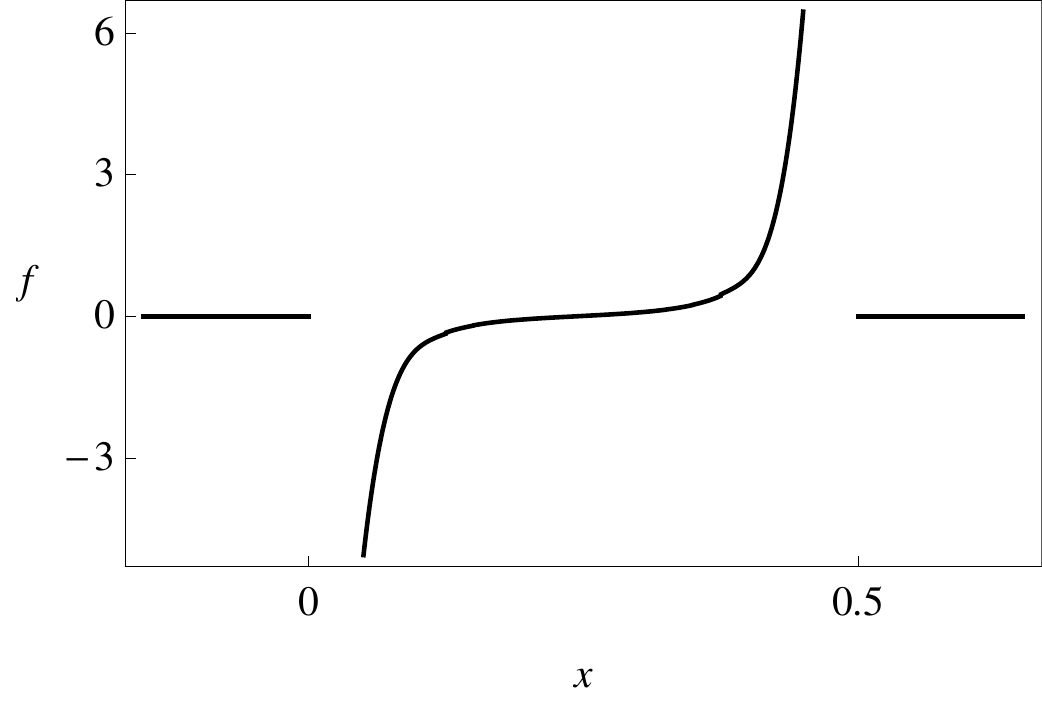}
\caption{The Casimir stress (\ref{xx1}) (left) and Casimir force per unit volume (\ref{f2}) (right) for the permittivity in Fig.~\ref{fig3} with $\hbar=c=1$, obtained using the regularization given by by (\ref{greg}) and (\ref{g0E})--(\ref{g0M}). The dots in the plot of $\sigma_{xx}$ show the results of numerical evaluation of the double integral in (\ref{xx1}) at those values of $x$; the line through the dots is a polynomial fit. In the homogeneous regions $x<0$ and $x>0.5$ the stress $\sigma_{xx}$ is zero. The force per unit volume $f$ in the inhomogeneous medium was obtained by differentiating the continuous fit to  $\sigma_{xx}$.}   \label{fig9}
\end{figure}

We now show the results for the alternative regularization (\ref{greg}) based on (\ref{g0E}) and (\ref{g0M}). Figures~\ref{fig7} and~\ref{fig8} show plots of the integrand $\widetilde{\sigma}_{xx}$ as a function of $u$ with $\xi$ fixed, and as a function of $\xi$ with $u$ fixed, for several values of $x$. It can be seen how the area under the curves in the $u$-plots increases significantly as $x$ moves towards $0$ or $x_R=0.5$, but the curves approach zero as $u\rightarrow\infty$ and $\xi\rightarrow\infty$ for $0<x<x_R$, giving a finite answer for the stress $\sigma_{xx}$ at these values of $x$. The behaviour of $\widetilde{\sigma}_{xx}$ as a function of $u$ when $x=0$ and $x=x_R$, however, shows a divergence of the stress $\sigma_{xx}$ at these positions, since $\widetilde{\sigma}_{xx}$ increases to a constant positive value as $u\rightarrow\infty$, giving an infinite value for the integral (\ref{xx1}). Figure~\ref{fig9} shows the resulting Casimir stress $\sigma_{xx}$, obtained by numerically integrating (\ref{xx2}) in (\ref{xx1}). The integrations must be performed separately for each $x$ and the dots show $\sigma_{xx}$ for a discrete set of $x$ values, the results for which are joined together by a polynomial fit. In the homogeneous material at $x<0$ and $x>0.5$ the stress $\sigma_{xx}$ is identically zero, as we verified by calculating the integrand (\ref{xx2}) in these regions where there is no difficulty with the standard regularization. Figure~\ref{fig9} also shows the corresponding Casimir force per unit volume (\ref{f2}); in the inhomogeneous region $f$ was found by differentiating the polynomial fit to  $\sigma_{xx}$.  The unlimited increase in $\sigma_{xx}$ as the interfaces  $x=0$ and $x=x_R$ are approached can be seen in  Figure~\ref{fig9}, leading to similar divergences in $f$.

The expectation of a u-shaped $\sigma_{xx}$, giving a Casimir force $f$ that changes direction as one moves across the inhomogeneous medium, is to some extent borne out by Fig~\ref{fig9}. It was the failure of the standard regularization recipe to give a finite stress and force per unit volume that led us to try the alternative proposal based on (\ref{g0E}) and (\ref{g0M}), an approach that at least has some \textit{a priori} physical justification. Although the alternative regularization gives a finite Casimir stress inside the inhomogeneous medium, the stress and the resulting force per unit volume are still infinite on the interfaces with the homogeneous regions. We have therefore not obtained a physically acceptable solution to the problem considered, with either regularization.

An interesting parameter value to investigate is $\varepsilon_R=1$, so that the inhomogeneous medium in Fig.~\ref{fig2} joins continuously to vacuum on the right-hand side. An example of this is plotted in Fig.~\ref{fig10}, where the other parameters are chosen as $\varepsilon_L=3$ and $b=1$, giving $x_R=\ln(3)$. The results for the stress $\sigma_{xx}$ and force per unit volume $f$ in this case are qualitatively similar to those for the previous parameter set. Figure~\ref{fig11} shows $\sigma_{xx}$ and $f$ computed using the alternative regularization based on (\ref{g0E}) and (\ref{g0M}). The Casimir force again diverges on the boundaries between the inhomogeneous medium and the homogeneous regions, even though the density of the inhomogeneous dielectric goes to zero at the right-hand boundary.
\begin{figure}[t]
\begin{center}
\includegraphics[width=18.0pc]{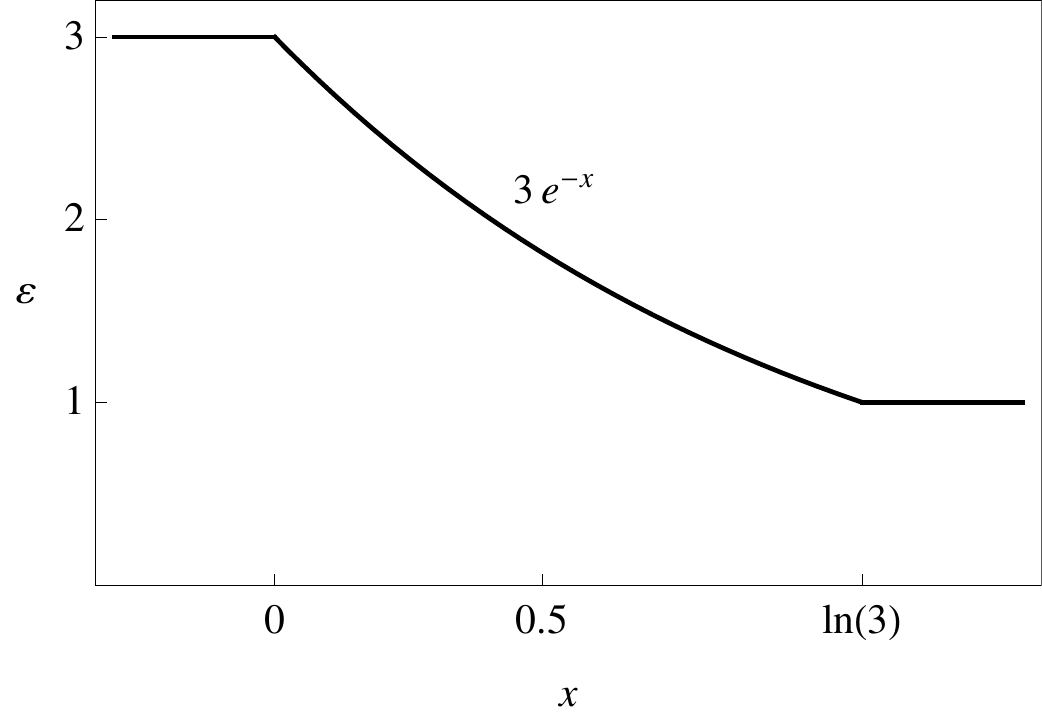}
\end{center}
\caption{The permittivity of Fig.~\ref{fig2} for the parameter values $\varepsilon_L=3$, $\varepsilon_R=1$ and $b=1$. This gives $x_R=\ln(3)$. In this case the homogensous medium on the right-hand side $x>x_R$ is vacuum.}   \label{fig10}
\end{figure}
\begin{figure}[t]
\includegraphics[width=18.0pc]{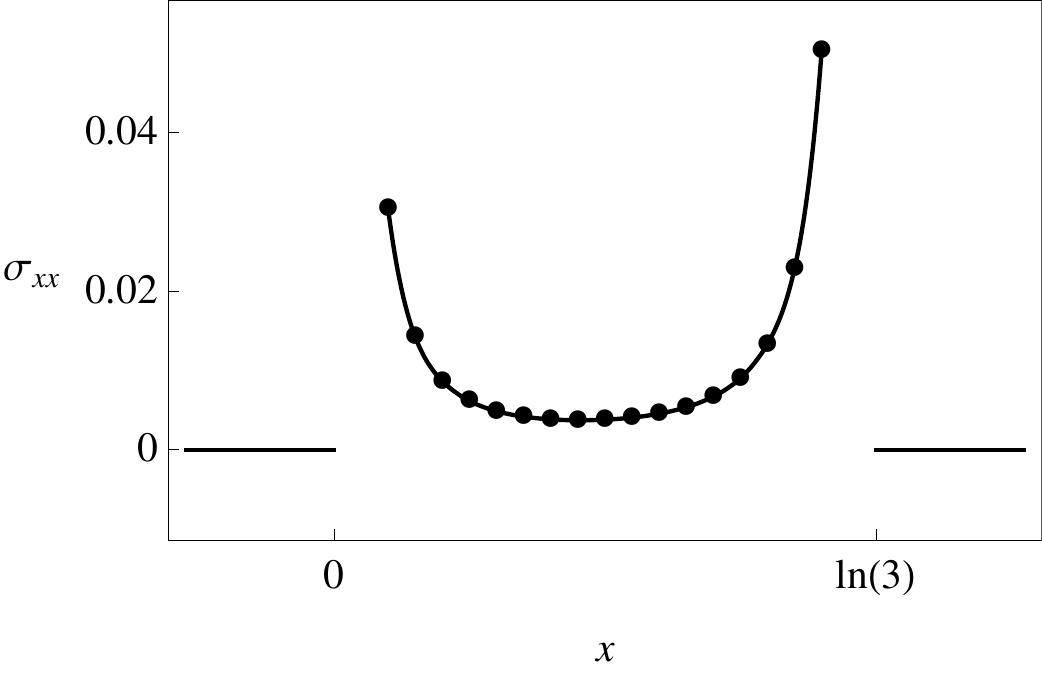}
\hspace{5mm}
\includegraphics[width=18.0pc]{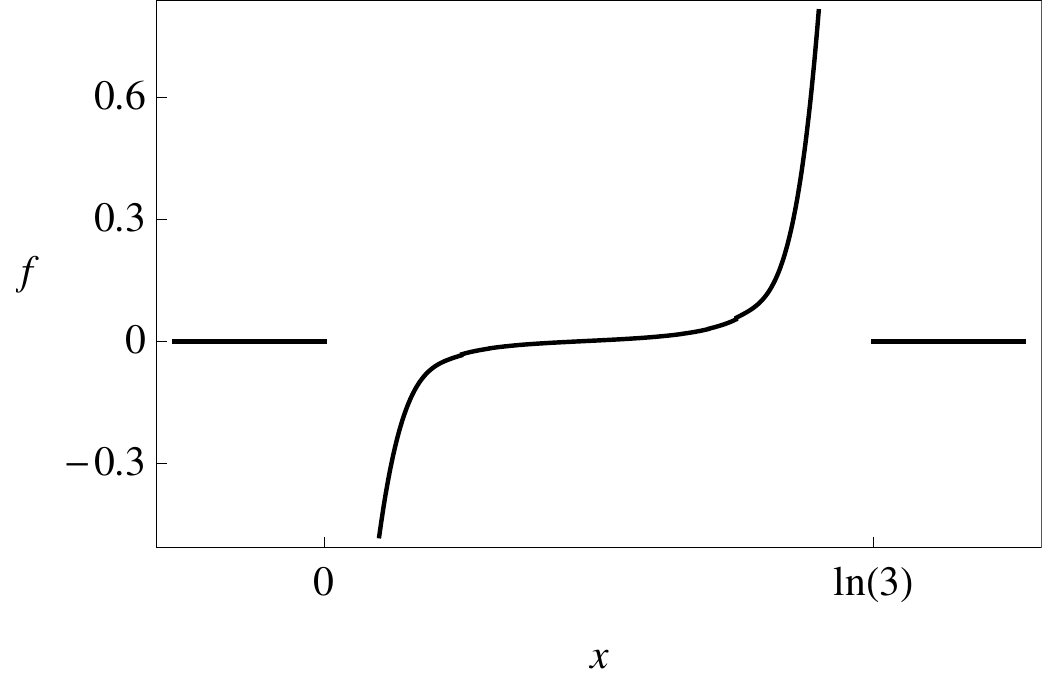}
\caption{The Casimir stress (\ref{xx1}) (left) and Casimir force per unit volume (\ref{f2}) (right) for the permittivity in Fig.~\ref{fig10} with $\hbar=c=1$, obtained using the regularization given by by (\ref{greg}) and (\ref{g0E})--(\ref{g0M}). As in Fig.~\ref{fig9}, the dots in the plot of $\sigma_{xx}$ show the results of numerical evaluation of the double integral in (\ref{xx1}). The force per unit volume $f$ in the inhomogeneous medium was obtained by differentiating the continuous fit to  $\sigma_{xx}$.}   \label{fig11}
\end{figure}

\section{Discussion}
We have presented a calculation of the Casimir self-force in an inhomogeneous dielectric, using a simple model for the dielectric permittivity. As far as we have been able to ascertain, this is the first analysis of the Casimir effect for inhomogeneous media, as opposed to piece-wise homogeneous media. We used the formalism of Lifshitz, which is routinely recognized as the most versatile and general theory of Casimir forces. Although the standard Lifshitz regularization prescription was formulated with the general case of inhomogeneous dielectrics in mind~\cite{dzy61,LL}, it gives a infinite Casimir stress everywhere inside the inhomogeneous medium in our example. An attempt to extract a finite Casimir force per unit volume from the diverging stress failed. We also tried a new regularization method, which aims to remove the contribution to the Casimir force arising from the inhomogeneity over short length scales where the use of macroscopic electromagnetism is unphysical. The new regularization gives a finite Casimir stress inside the inhomogeneous medium in the example considered, but the stress and force per unit volume increase without limit at the boundaries joining the inhomogeneous dielectric to homogeneous regions. Thus, even with the new regularization, we have been unable to solve the example using the accepted theoretical machinery. 

It is probable that if the sharp corners in $\varepsilon$ at $x=0$ and $x=x_R$ in Fig.~\ref{fig2} are smoothed out sufficiently then the new regularization procedure will give a finite Casimir stress $\sigma_{xx}$ everywhere. This certainly seems to be the case for a permittivity profile given by (\ref{perm}) for \textit{all} $x$, although we do not discuss this example here since $\varepsilon\rightarrow\infty$ for $x\rightarrow-\infty$ and $\varepsilon\rightarrow 0$ for $x\rightarrow\infty$ is not a very instructive model. In this regard it is important to reflect on the fact that the new regularization method achieved a finite stress inside the inhomogeneous region by removing the contribution of small-scale inhomogeneity. The sharp corners in Fig.~\ref{fig2} are clearly at variance with an approach that views small-scale inhomogeneity in $\varepsilon$ as unphysical and the regularization procedure does not apply at the corners themselves since they are boundaries between regions where the stress is regularized separately. A model where the function $\varepsilon(x)$ is $C^n$ ($n\geq 1$) everywhere and constant for $x<0$ and $x>x_R$ would probably have to be solved entirely numerically. Another case of interest would be a function $\varepsilon(x)$ with discontinuities where $d\varepsilon(x)/dx\rightarrow 0$ at each side of the jump, as in Fig.~\ref{fig1}; it is unclear if the new regularization would always give a finite stress in these circumstances.

On the other hand, it is conceivable that the divergence of the  
regularized Casimir force is not an artifact of the theory. The  
Casimir force could indeed tend to infinity at discontinuities in the  
derivatives of the dielectric profiles, for example at the sharp  
corners in the graph of the dielectric function of Fig. 2. In response  
to such a strong force, the dielectric may either consolidate a sharp  
boundary between dielectric layers, where the regularized Casimir  
force is finite, or smooth out the transition region. But of course,  
these ideas are speculations at present.

As discussed earlier in this paper, it has been known for some time that Lifshitz theory also founders for general spherically or cylindrically symmetric dielectrics, and the Casimir force in these cases is unknown. A legitimate line of enquiry is whether a Casimir self-force on a sphere or cylinder, or on an inhomogeneous dielectric, is measurable. This consideration may play a part in resolving the difficulties, but such materials certainly alter the local quantum vacuum and this affects the total stress-energy-momentum of the system. The total energy-momentum tensor of a material has a local significance as the source of the gravitational field, so it needs to be explained exactly how this quantity is to be computed, for any material, and this requires an understanding of the Casimir effect that is currently lacking. Lifshitz theory takes a semi-classical approach to the Casimir effect, wherein the electromagnetic field is quantized but the matter is treated classically. It may be that the regularization difficulties that have been encountered here and elsewhere are due to an inherent limitation of this semi-classical approach. Alternatively, a better understanding of the physical basis of the regularization of quantum-vacuum energy may resolve the issue, allowing a prediction of the Casimir force in all situations where the required Green functions are known. A final, encouraging thought is that experimental investigation of Casimir forces is a maturing discipline~\cite{bor09}, so there is reason to hope that experiments can be designed that will shed some light on these areas where the theory is sorely underdeveloped.

\section*{Acknowledgements}
We are grateful to Tom Kelsey and Steve Linton for their continuing  
encouragement and generous help. Our research is supported by the  
Royal Society of Edinburgh, the Scottish Government, the Centre for  
Interdisciplinary Research in Computational Algebra, EPSRC grant EP/CS23229/1, and a Royal  
Society Wolfson Research Merit Award.


\section*{Author contributions}
The problem of the Casimir stress in an inhomogeneous medium was  
suggested by UL. UL and TP developed the general expression for the  
Green tensor in 1D-inhomogeneous dielectrics. TP and CX investigated  
simple models and the specific case solved in the paper was chosen by  
TP. TP and CX independently computed the Casimir stress, using  
different software packages. The derivation of the analytic solutions  
for the Green functions was prepared by TP based on the computer  
solutions by CX and TP. CX found the intrinsic constant problem in the  
Lifshitz regularization. The new regularization method was developed  
by UL. The results for the Casimir force per unit area using the  
conventional regularization were obtained by TP. The figures and plots  
in the paper were prepared by TP, who wrote the text, with an addition  
by UL.

\end{document}